\documentclass[twocolumn,useAMS,usenatbib]{mn2e}
\usepackage{natbib}
\usepackage{amsmath}
\usepackage{latexsym}
\usepackage{epsfig,graphicx}
\usepackage{subfig}
\usepackage{float}
\usepackage{color}
\usepackage{hyperref}
\graphicspath{{Figures/}}
\newcommand{\new}[1]{#1}

\def\reff@jnl#1{{\rm#1\/}}

\def\aj{\reff@jnl{AJ}}                  
\def\araa{\reff@jnl{ARA\&A}}            
\def\apj{\reff@jnl{ApJ}}                
\def\apjl{\reff@jnl{ApJ}}               
\def\apjs{\reff@jnl{ApJS}}              
\def\apss{\reff@jnl{Ap\&SS}}            
\def\aap{\reff@jnl{A\&A}}               
\def\aapr{\reff@jnl{A\&A~Rev.}}         
\def\aaps{\reff@jnl{A\&AS}}             
\def\baas{\reff@jnl{BAAS}}              
\def\jrasc{\reff@jnl{JRASC}}            
\def\memras{\reff@jnl{MmRAS}}           
\def\mnras{\reff@jnl{MNRAS}}            
\def\physrep{\reff@jnl{Phys.Rep.}}
\def\pra{\reff@jnl{Phys.Rev.A}}         
\def\prb{\reff@jnl{Phys.Rev.B}}         
\def\prc{\reff@jnl{Phys.Rev.C}}         
\def\prd{\reff@jnl{Phys.Rev.D}}         
\def\prl{\reff@jnl{Phys.Rev.Lett}}      
\def\pasp{\reff@jnl{PASP}}              
\def\pasj{\reff@jnl{PASJ}}              
\def\skytel{\reff@jnl{S\&T}}            
\def\solphys{\reff@jnl{Solar~Phys.}}    
\def\sovast{\reff@jnl{Soviet~Ast.}}     
\def\ssr{\reff@jnl{Space~Sci.Rev.}}     
\def\nat{\reff@jnl{Nature}}             

\newcommand{\beq}{\begin{equation}}
\newcommand{\eeq}{\end{equation}}
\newcommand{\beqa}{\begin{eqnarray}}
\newcommand{\eeqa}{\end{eqnarray}}

\newcommand{\sersic}{S\'{e}rsic }

\newcommand{\s}{\ensuremath{\mathcal{S}}}
\newcommand{\scinot}[2]{\ensuremath{#1 \times 10^{#2}}} 

\title[Cosmic variance in simulations]{The impact of cosmic variance on simulating weak lensing surveys}

\author[Kannawadi et al.]
{Arun Kannawadi$^1$\thanks{\tt akannawa@andrew.cmu.edu}, 
Rachel Mandelbaum$^1$,
Claire Lackner$^2$
\\$^1$McWilliams Center for Cosmology, Carnegie Mellon University, Pittsburgh, PA 15217, USA
\\$^2$Kavli Institute for the Physics and Mathematics of the Universe (WPI), Todai Institutes for Advanced Study,\\ the University of Tokyo, Kashiwa, Japan.
}

\date{\today}

\begin{document}
\bibliographystyle{mn2e2}
\maketitle

\begin{abstract}
Upcoming weak lensing surveys will survey large cosmological volumes to measure the growth of cosmological structure with time and
thereby constrain dark energy.  One major systematic uncertainty in
this process is the calibration of the weak lensing shape distortions,
or shears.  Most upcoming surveys plan to test several aspects of their shear
estimation algorithms using sophisticated image simulations that
include 
realistic galaxy populations based on high-resolution data from the
Hubble Space Telescope ({\em HST}).  However, existing datasets
from the {\em HST} cover very small cosmological volumes, so cosmic variance could cause the galaxy populations in them
to be atypical. A narrow redshift slice from such surveys could be dominated by a single large overdensity or underdensity.  In that case, the
morphology-density relation could alter the local galaxy
populations and yield an incorrect calibration of shear estimates as a function of redshift. We directly test this scenario using the COSMOS survey, the largest-area {\em HST} survey to date, and show how the
statistical distributions of galaxy shapes and
morphological parameters (e.g., S\'{e}rsic $n$)
are influenced by redshift-dependent cosmic variance. 
The typical variation in RMS ellipticity due to environmental effects
is 5 per cent (absolute, not relative) for redshift bins of width $\Delta z=0.05$, which could result in uncertain shear calibration at the 1 per cent level. 
We conclude that the cosmic variance effects are large enough to exceed the systematic error budget of future surveys, but can be mitigated with careful choice of training dataset and sufficiently large redshift binning.
\end{abstract}


\section{Introduction}
\label{S:intro}

Weak gravitational lensing, the deflection of light by mass, is one of the cleanest ways to study the nature of dark energy by tracking the growth of structure in the Universe as a function of time \citep[e.g.,][]{2001PhR...340..291B,2006astro.ph..9591A,2013PhR...530...87W}.
As light from background sources passes by matter (including dark matter) on its way to us, the apparent shapes of the background galaxies get distorted, and the galaxies get slightly magnified as well.
Because of its sensitivity to dark matter and dark energy, major surveys such as the Hyper Suprime-Cam \citep[HSC;][]{HSC_Overview}, Dark Energy Survey~\citep[DES;][]{DESC}, the KIlo-Degree Survey~\citep[KIDS;][]{KIDS}, the Panoramic Survey Telescope and Rapid Response System~\citep[PanSTARRS;][]{PanSTARRS_2010},
the Large Synoptic Survey Telescope~\citep[LSST;][]{LSST_Book}, Euclid\footnote{\url{http://sci.esa.int/euclid/}, \url{http://www.euclid-ec.org}} \citep{EuclidReport},
and Wide-Field Infrared Survey Telescope~\citep[WFIRST;][]{WFIRST_Final}
are planned for the next two decades to gather enormous quantities of weak lensing data that will lead to precise constraints on the growth of structure with time, and therefore cosmological parameters.

For the upcoming surveys to achieve their promise, their systematic error budgets must be below their statistical error budgets.
Systematic error budgets for weak lensing surveys typically include
astrophysical effects, such as intrinsic alignments of galaxy shapes
with large scale density fields \citep[e.g.,][]{2014arXiv1407.6990T} and the effect of baryons on the
matter power spectrum \citep[e.g.,][]{2011MNRAS.415.3649V,2011MNRAS.417.2020S}, as well as observational uncertainties such as
the ability to robustly infer shears from observed galaxy shapes or
photometric redshifts from their observed colours.
Given the expected sub-per cent statistical errors on upcoming surveys, systematic errors must be reduced from
 their typical level in the current state-of-the-art
 measurements that typically achieve $\sim 5$ per cent statistical
 errors at best
 \citep[e.g.,][]{2010A&A...516A..63S,2013MNRAS.432.2433H,2013ApJ...765...74J,2013MNRAS.432.1544M}.

Some types of information about and tests of shear estimation
   for ongoing and future weak lensing surveys will rely on data from
   the Hubble Space Telescope ({\em HST}). 
    For example, data from
   {\em HST} can be used to derive basic statistics of galaxy light
   profiles, such as joint distributions of size and the morphology.
   It can be used to infer the intrinsic distribution of galaxy
   shapes, which enters the shear estimation process either implicitly
 or explicitly depending on the method used for shear estimation
 \citep[see, e.g.,][]{2014MNRAS.439.1909V}.
 Going beyond basic information about the light profile, {\em HST} can
be used to quantify the detailed morphology of galaxies, due to its
higher resolution compared to any current or planned weak lensing
survey.  Finally, it can be used to address systematics due to colour
gradients within galaxies, which are particularly problematic when
combined with a wavelength-dependent diffraction-limited PSF
\citep[e.g.,][]{2012MNRAS.421.1385V,2013MNRAS.432.2385S}.

One method that is commonly used to test for the presence of
systematic errors in the shear estimation process is image simulation,
where we can cleanly test whether our methods of shear estimation
recover the ground truth. This is a valuable test, considering the
numerous sources of additive and multiplicative bias such as a
mismatch between galaxy model assumptions and actual galaxy light
profiles \citep[e.g.,][]{2010MNRAS.404..458V, 2010A&A...510A..75M}, biases due to the effects of pixel noise on the
shear estimates
\citep{2012MNRAS.427.2711K,2012MNRAS.424.2757M,2012MNRAS.425.1951R},
and ellipticity gradients \citep{2010MNRAS.406.2793B}.  These biases
often differ for galaxies with different morphologies (e.g., disks
vs.\ ellipticals), sizes, $S/N$, and shape \citep{2010MNRAS.405.2044B,2012MNRAS.423.3163K}. 
A general requirement for simulations used to test shear recovery is
that they should be as realistic as possible.
 
 Realistic simulations may use samples based on images from the 
 ({\em HST}). Software packages like
{\sc GalSim}\footnote{\url{https://github.com/GalSim-developers/GalSim}}
\citep{2014arXiv1407.7676R} can generate images of galaxies from the
{\em HST} as they would appear with an additional lensing shear and
viewed by some lower resolution telescope.  Examples of training samples from the {\em HST} include 
the COSMOS survey \citep[used by the GREAT3 challenge,][]{great3} or the
 Ultra Deep Field \citep[UDF, used by][]{2013ApJ...765...74J}.  These
 two examples serve as the extremes in the {\em HST} samples used as
 the basis for image simulation, with COSMOS being 
the widest contiguous area surveyed by the {\em HST} currently and hence representive, and the 
latter being extremely deep but narrow.

For a variety of physical reasons, some of which are still not fully
understood, the shape and morphology of
galaxies depends on their local environment
\citep[e.g.,][]{2014arXiv1402.1172C,2014MNRAS.444.2200D}. 
Hence, local overdensities or underdensities along the line of sight
observed in these {\em HST} fields may (given the small size of the field)
cause the properties of the galaxy population in redshift slices to be
atypical depending on the environment in that slice.  This has
the undesired consequence of including a variation in galaxy properties
due to the COSMOS (or other) survey cosmic variance in the simulated galaxy sample in that redshift slice, rather than
only including ensemble effects that would appear in a large cosmological volume, such as true redshift evolution of
galaxy properties.  Our goal is to quantify the degree to which the
morphology-density correlations in COSMOS cause noticeable changes in
the galaxy populations in narrow redshift slices at a level that could
result in difficulty using the sample to derive redshift-dependent
shear calibrations.   Upcoming surveys will study lensing as a function of
redshift and therefore need to simulate galaxy samples at different
redshifts in order to assess the shear calibration at each redshift.

The paper is structured as follows: in Sec.~\ref{S:data}, we describe
the data that we use for this study.  In Sec.~\ref{S:methods}, we
describe our methods for deriving the relevant galaxy properties
like environment, morphology, and shape.  
Using these ingredients, we present our results in Sec.~\ref{S:results}
and discuss their implications in Sec.~\ref{S:implications},
concluding in Sec.~\ref{S:summary}.
\section{Data}
\label{S:data}
The COSMOS survey \citep{COSMOS_overview, COSMOS_generic, COSMOS_Alexie} is a flux-limited, narrow deep field survey covering a contiguous area of $1.64 \text{ deg}^2$ of sky, with images taken using the Advanced Camera for Surveys (ACS) Wide Field Channel (WFC)
in the Hubble Space Telescope (HST).  We use the COSMOS survey to
define a parent sample of galaxy images to be used for making image
simulations, following the approach taken in
\cite{2012MNRAS.420.1518M,great3}.

We apply the following set of initial cuts to the COSMOS data, the
first two of which
are motivated and explained in more detail by \cite{COSMOS_Alexie}:
\begin{enumerate}
 \item \texttt{MU\_CLASS=1}: This criterion uses a comparison between
   the peak surface brightness and the background level to achieve a
   robust star/galaxy separation, with galaxies having \texttt{MU\_CLASS=1}. 
 \item \texttt{CLEAN=1}: Objects near bright stars or those containing
   saturated pixels were removed; the rest pass this cut on \texttt{CLEAN}. 
 \item \texttt{GOOD\_ZPHOT\_SOURCE =1}: This cut requires that
   photometric redshifts be reliable and good enough to draw
   conclusions about the population (see \citealt{2012MNRAS.420.1518M}
   for details).
\end{enumerate}

High resolution images taken through the wide F814W filter (broad
\emph{I}) for all galaxies passing the above cuts were used to create
a collection of postage stamp images for the GREAT3 challenge
\citep{great3}, using the procedure described in
\cite{2012MNRAS.420.1518M}.  Each galaxy postage stamp image has a
corresponding PSF image that can be used by {\sc GalSim} or other
software to remove the effects of the {\em HST} PSF before simulating
the galaxy image as it would appear at lower resolution.

To better characterize the galaxy population, parametric models were
fit to the light profiles of these galaxies.  These were carried out
using the method described in \cite{Claire_Fits}, and include
\sersic profile fits and 
2 component bulge $+$ disk fits described in detail in \cite{great3}
and briefly in Sec.~\ref{sub:axisratio} of this work.  

In addition to the ACS/WFC (F814W) imaging, the COSMOS field has also been imaged by Subaru Suprime-Cam, the
Canada-French Hawaii Telescope (CFHT) and KPNO/CTIO, yielding many bands
of imaging data used to determine high-fidelity photometric
redshifts. 
Photometric redshifts were determined by
\cite{COSMOS_Photoz_30band}. The accuracy of photometric redshifts for
$m_\text{F814W}\le$ 22.5 is $\sigma_{\Delta z} = 0.007(1+z)$; for $m_\text{F814W}\le$ 24, $\sigma_{\Delta z} = 0.012(1+z)$.
The photometric redshift values become noisier beyond $z\sim 1.2$, and
the  fits to the galaxy light profiles are also somewhat noisy once we
go beyond $m_\text{F814W}\sim 23.5$.   
For this reason, we will exclude all galaxies that have F814W
magnitude fainter than 23.5. However, we will use the
$m_\text{F814W}\le 25.2$ sample that was generated for the GREAT3
challenge to estimate the completeness, which is useful when
generating a volume-limited sample (Sec.~\ref{sub:volumelimiting}).
We first use the $z\le1.25$ flux-limited sample to fit parametric redshift
distribution models (Sec.~\ref{sub:overdensities}), and then
restrict ourselves to $z\le1$ sample for all further analysis. 

Stellar mass estimates were obtained~\citep{COSMOS_XRAY} using the
Bayesian code described in~\cite{KEVIN_MSTAR}.  This process involves
constructing a grid of models that vary in age, star formation
history, dust content and metallicity (always assuming a Chabrier IMF;
\citealt{ChabrierIMF}), 
 to which the observed galaxy spectral energy distributions (SEDs) and photometric redshift are compared. At each grid point, the probability that the 
SED fits the model is calculated, and by marginalizing over the nuisance parameters in the grid, the stellar mass probability distribution is obtained. The median of this distribution
is taken as the stellar mass estimate. 
  
\section{Methods}
\label{S:methods}

In order to study the variation in the intrinsic ellipticity
distribution and various morphological indicators with the galaxy environment, there are three main steps to be carried out:
\begin{enumerate}
 \item Identify overdense and underdense environments along the line of sight in our survey
   from the redshift distribution of galaxies (Sec.~\ref{sub:overdensities});
 \item volume-limit the sample such that Malmquist bias is minimized
   before comparing galaxies in different redshift slices
   (Sec.~\ref{sub:volumelimiting}); and 
 \item estimate the galaxy axis ratios and other morphological indicators such as  \sersic index and bulge-to-total ratios (Sec.~\ref{sub:axisratio}).
\end{enumerate}

In this section we will describe how these steps were carried out.

\subsection{Finding overdensities}
\label{sub:overdensities}
\begin{figure}
 \centering
  \includegraphics[width=\columnwidth]{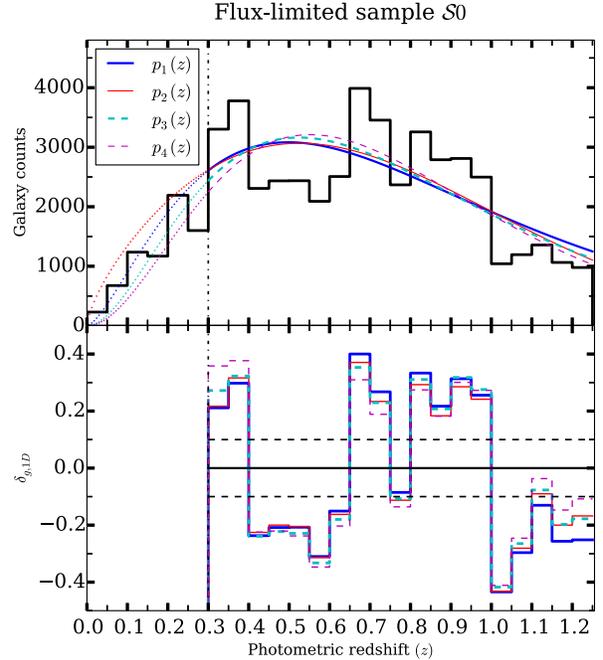}
  \caption{Upper panel: Redshift distribution of flux-limited
    ($m_\text{F814W}\le 23.5$) sample with photometric redshift bins
    that are 0.05 wide. The vertical line at $z=0.3$ indicates the delineation between lower redshifts that we do not use for our analysis, and higher redshifts that are used. Fits to two analytical functions, $p_1(z)$ and $p_2(z)$, defined in Eqs.~\ref{E:pz1} and~\ref{E:pz2}, are also shown, with best-fitting parameters $a=2.53\pm0.98$, $z_1=0.32\pm0.16$, $b=1.70\pm0.50$ and $z_2=0.63\pm0.13$.  We also show the distributions from \protect\cite{2004ApJ...617..765C} from Eqs.~\ref{E;pz3} and~\ref{E;pz4}. 
           Lower panel: Plot of $\delta_{g,\text{1D}} =
           N/N_{\text{mod}}-1$ with each functional form as the model
           for each redshift bin.
}
  \label{fig:redshift_fluxlimited}
\end{figure}

It is important to keep in mind when considering the environment
estimation that our goal is not to create a full 3D mapping of the
density field within the COSMOS region (a task that was already
addressed by \citealt{Kovac_Density10k} using the zCOSMOS
spectroscopic sample).  Instead, we make a coarse, 1-dimensional, line of sight division of the
COSMOS survey into redshift slices, just as would be done when making
galaxy redshift slices as input to a weak lensing survey simulation.
For each redshift slice, we can then check whether the environment is
overdense or underdense on average.  Our approach will tend to wash
out some real trends from a 3D study, but is appropriate given our
scientific goal of testing effects of the environment on weak lensing
simulations based on the COSMOS survey.

For our (flux-limited) sample of galaxies, up to $z=1.25$, we fit
parametric models to the histograms of photometric redshifts in order
to assign values of overdensity.  We choose our bins to be $0.05$ wide
starting from $z=0.3$, where the bin width is selected to be somewhat
larger than the photometric redshift error but narrow enough that we
can still identify rather than averaging over real cosmological
structures.  We neglect the lowest
redshifts which have negligible cosmological volume and where the
galaxy population tends to be intrinsically bright and large enough
that a non-negligible fraction is lost due to the cuts we impose
(Sec.~\ref{S:data}).

The parametric redshift distributions that we use are
\begin{equation}\label{E:pz1}
 p_1(z) \propto z^{a-1}\exp\left[{-z/z_1}\right]
\end{equation}
and 
\begin{equation}\label{E:pz2}
 p_2(z) \propto z^{b-1}\exp\left[ -\frac{1}{2}\left( \frac{z}{z_2} \right)^2\right]
\end{equation}

Here $a$, $b$, $z_1$ and
$z_2$ are free parameters that are to be determined.  The
normalization constants depend not only on the parameters but also on
the lower and the upper limit of the redshifts considered, where we
fix the normalization to ensure that the predicted number of galaxies
in the range used ($0.3<z<1.25$) is equal to the actual number. 
Fig.~\ref{fig:redshift_fluxlimited} shows the photometric redshift
histogram together with the best-fitting parametric distributions.

We compare our fits to the fits made by \cite{2004ApJ...617..765C} using 
the DEEP2 galaxy redshift survey. Interpolating between their results for the $18<I_\text{AB}<23$
and $18<I_\text{AB}<24$ samples to our own limiting magnitude (and assuming equivalence of our $I$ bands), we obtain the following redshift distributions:
\begin{equation}
 \label{E;pz3}
 p_3(z) \propto z^2 \exp{[-z/0.262]}
\end{equation}
and 
\begin{equation}
 \label{E;pz4}
 p_4(z) \propto z^2 \exp{[-(z/0.361)^{1.2}]}
\end{equation}
which are also plotted in Fig.~\ref{fig:redshift_fluxlimited}.  
Note that $p_3(z)$ is a special case of $p_1(z)$ with $a=3$ and
$z_1=0.262$; these parameter values are within the $1\sigma$ allowed
regions for our fits to Eq.~\ref{E:pz1}.  Visually, these
distributions appear quite similar to our own fits carried out here, which is
reassuring given the use of different survey data and functional forms.

The estimated overdensity in a redshift bin is defined by comparing
the observed galaxy counts in the bin with the counts that are
predicted in that bin by one of the models in Eqs.~\eqref{E:pz1}
and~\eqref{E:pz2}:
\begin{equation}
\delta_{g,\text{1D}}=\frac{(N-N_{\text{mod}})}{N_{\text{mod}}},
\end{equation}
where
\begin{equation}
N_\text{mod} = \int_{z_\text{min}}^{z_\text{max}} p(z) \,\mathrm{d}z
\end{equation}
is determined by integrating the redshift distribution within the
limits of that redshift slice. 
Note that $\delta_{g,\text{1D}}$ is dependent on our choice of model
redshift distribution, and should have a mean value of $0$ over the
entire redshift range when weighted by the number fraction in each bin.

Our preliminary decision criterion for identifying overdense and underdense
redshift slices involves leaving a 10 per cent margin around an
overdensity of zero; i.e., if $\left| \delta_{g,\text{1D}} \right| <
0.1$, that is considered ``neutral'' (neither overdense nor
underdense on average).  We can then label each redshift slice as
either overdense, underdense, or neutral as follows: 
We label a redshift bin as overdense if at least one model gives a
value of $\delta_{g,\text{1D}} > 0.1$ while the other gives
$\delta_{g,\text{1D}} > -0.1$ 
(neutral or overdense), and vice versa for the underdense regions. We
label a redshift bin as neutral if both models give
$\delta_{g,\text{1D}}$ within the neutral region,  
\emph{or} if use of one model redshift distribution results in the
conclusion that the bin is overdense while the other leads to the
conclusion that it is underdense. 

Once we volume-limit our sample (explained in Sec.~\ref{sub:volumelimiting}), we again compare the histogram to the models.
If there is no qualitative change in the overdensities, we stick with the preliminary decision. If the overdensity values
flip in sign or become too small, then we classify the redshift bin as `neutral'.
The naive Poisson uncertainty of the counts
in each bin is much less than the difference between the actual number of galaxies present
and the number predicted by the models. For this reason, errorbars have been completely ignored.

We thus identify the regions $z=0.30-0.40$, $0.65-0.75$, and
$0.80-0.85$ as overdense; $z=0.55-0.65$ and $0.75-0.80$ as underdense;
and we defer classication of the somewhat ambiguous range from
$z=0.40-0.55$ until Sec.~\ref{sub:volumelimiting}. 

We have adopted this purely 1D environment classification for reasons
explained at the beginning of this section. However, as a sanity check
we can compare it with a more rigorous study that includes information
about structure in the plane of the sky.  
 \cite{Kovac_Density10k} used a sample of $\sim$10~000 zCOSMOS
 spectroscopic galaxies with $I_{AB}<22.5$ to reconstruct the three dimensional overdensity field up to $z\sim 1$.
We find that our classification of overdensities and underdensities
agrees with this work, except for our two highest redshift bins.
We believe that this disagreement is due to the errors in our
photometric redshifts, with the overdensity reported by
\cite{Kovac_Density10k} in the $z=0.875-1$ range 
leaking into our $z=0.80-0.85$ slice.

\begin{figure}
  \centering
   \includegraphics[width=\columnwidth]{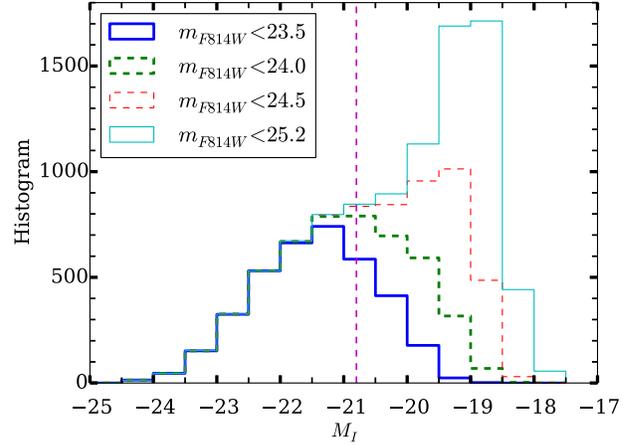}
   \caption{Distribution of absolute magnitude $M_I$ for various
     flux-limited samples in the redshift range $[0.80, 0.85]$ are plotted together. The vertical line
     corresponds to our luminosity cut of $-20.8$, brighter than which
     the $m_\text{F814W}<23.5$ sample includes $>95.3$ per cent of the
     galaxies in the $m_\text{F814W}<25.2$ sample.
   }
   \label{fig:MIhist}
 \end{figure}

\subsection{Volume-limiting}
\label{sub:volumelimiting}
 
COSMOS is a flux-limited survey and is therefore affected by Malmquist
bias, with the galaxy samples at higher redshifts being intrinsically
brighter on average.  
Our analysis involves comparing galaxies in different redshift slices
to identify significant differences in morphology that arise due to
morphology-density correlations.  Such an analysis would be very
difficult with a flux-limited sample because there would be some
variation in morphology with redshift just due to the intrinsic change
in the sample properties.  For a fair comparison, we must restrict
ourselves to galaxies that are bright enough that they would be
observed at all redshifts that we consider, which is achieved by
volume-limiting the sample.  We consider three different ways of
carrying out this process, which results in three different galaxy
sample selections, all of which we will use in the remainder of the												
analysis.

Call the flux-limited ($m_\text{F814W} \le 23.5$) sample \s$0$. Our first approach is to generate a volume-limited sample that is
complete up to $z=1$, by applying a cut on luminosity such that only
galaxies instrinsically brighter than a certain threshold (determined
in detail below) are considered. This threshold is set on the
$k$-corrected $I$-band absolute magnitudes ($M_I$) from the COSMOS PSF-matched
photometry catalog.   
Since the parent sample contains fainter galaxies and is quite
complete to $m_\text{F814W}=25.2$,
we compare the $M_I$ distribution
of the $m_\text{F814W}=23.5$ sample with flux-limited samples that
have fainter flux limits, to see where the $m_\text{F814W}<23.5$
sample that we want to use for our tests is no longer complete.
At $M_I\sim-22.0$, the
$m_\text{F814W}<23.5$ sample is beginning to lose galaxies in the
$0.9<z<1.0$ redshift bin due to the flux limit. However, the $0.85<z<1$
redshift bin was found to be only moderately overdense, so we choose
to disregard this region for the rest of the analysis, and instead
restrict to $z<0.85$, which is advantageous because it allows us to
choose a somewhat fainter intrinsic luminosity limit for the
analysis.  We relax our luminosity cut
so that the sample is volume-limited \emph{not} until $z=1$ but until $z=0.85$. 
We impose the cut at $M_I=-20.8$, which gives 95.3 per cent completeness in
the $0.8 < z \le 0.85$ bin (see Fig.~\ref{fig:MIhist}). The resulting
sample, which has 13~567 galaxies, will be called sample \s$1$ in the remainder
of this work.

However, previous studies
\citep[e.g.,][]{2003A&A...401...73W,2005ApJ...622..116G,2006ApJ...647..853W,Faber2007}
have shown that galaxy intrinsic luminosities evolve with
redshift. Thus, we should also let the luminosity cut that we apply to
volume-limit the sample evolve with redshift. 
Unfortunately, the majority of the published work on evolution of the
luminosity function uses $B$ and $V$ band data, and it is not apparent
that the results should be the same in a redder passband like $I$.  
We use the results from \cite{Faber2007} for the evolution of
$B$-band magnitudes from the DEEP2 and COMBO-17 surveys, which is $
\Delta M_B^* \sim -1.23$ mag per unit redshift (with the sign
indicating that galaxies were intrinsically brighter in the past), for a combined sample
of blue and red galaxies. 
Typically, estimates of evolution in the redder bands are less than the estimates of evolution in bluer bands
\citep{1999ApJ...518..533L, 2003ApJ...592..819B}.
Assuming that the evolution is a smooth function of the wavelength,
the evolution in $I$-band should be in between $B$ and $K$ band. 
Therefore, by considering no evolution (a lower limit) as in our \s$1$, and a second
sample \s$2$ constructed using the $B$-band evolution (as an upper
bound on the $I$-band evolution), we can assume that these two samples
bracket reality.

Thus, \s$2$ is constructed by letting the luminosity cut evolve,
starting from $M_I = -20.8$ (same as in \s$1$) for the $0.8<z\le0.85$
bin.  The cut values for the other bins are defined by allowing $1.23$
magnitudes of evolution to fainter magnitudes as a function of
redshift (evaluated using the bin centers).  Because of the sign of
redshift evolution, \s$2$ includes  more galaxies (15~903 galaxies). 

One might wonder why we cannot use the luminosity function in F814W
based on the COSMOS observations to directly determine the rate of evolution of the luminosity
function for our sample, thus simplifying this exercise.  However,
this turns out to be highly non-trivial for two reasons.  First, the
F814W observations are relatively shallow compared to the deep
ground-based observations used in many other works for determination
of luminosity evolution.  As a result, it is difficult to get a handle
on the faint end of the luminosity function, and the unknown faint-end
slope turns out to be degenerate with the evolution of the typical
luminosity.  Second, the photometric redshift error is a complicating
factor that requires sophisticated techniques to remove.  A derivation
of the $I$-band luminosity evolution is therefore beyond the scope of
this work.

\begin{figure*}
 \centering
 \includegraphics[width=2.2\columnwidth]{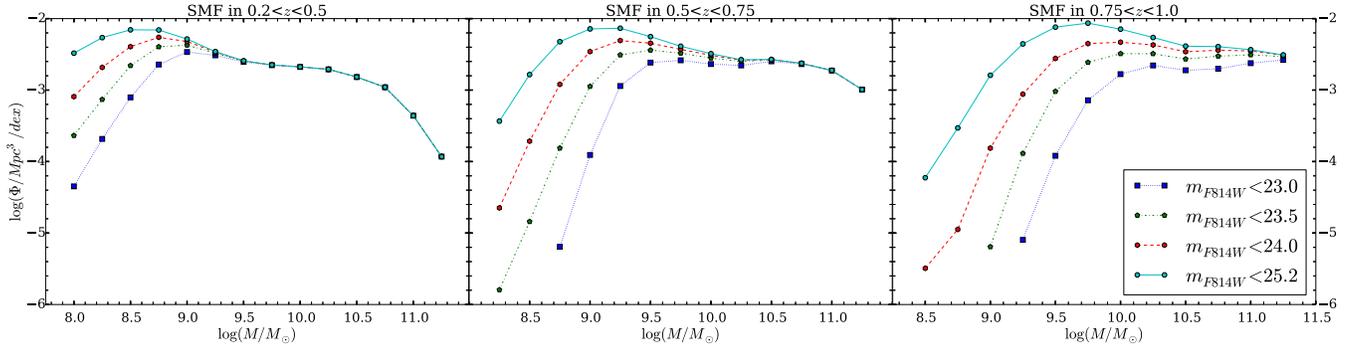}
 \caption{Stellar mass distribution for various flux-limited samples
   are shown in three redshift ranges as separate panels. The redshift
   bins have been chosen to facilitate the comparison with  
          a study of the SMF in \protect\cite{Tomczak_SMF}. At high mass, the
          distributions are the same for various flux limits,
          indicating that the samples are complete in that mass
          range. The curves begin to deviate at low masses due to
          incompleteness coming from the flux limit.  The point at
          which the deviation exceeds our threshold determines where the mass cutoff should be to volume-limit the sample.
          }
 \label{fig:smf}
\end{figure*}

Finally, we can circumvent the problem of redshift evolution of the
luminosity by imposing cuts on stellar mass instead. In Fig.~\ref{fig:smf}, we show the stellar mass function (SMF) of our sample for various F814W flux limits.
The shapes of the SMF curves we obtain are consistent with those in \cite{Tomczak_SMF} at the high stellar-mass end.
\cite{Tomczak_SMF} report the SMFs for the ZFOURGE survey, which
includes COSMOS. They calculated stellar masses using the procedure
and software described by \cite{2009ApJ...700..221K}, using a set of models
with exponentially declining star formation history \citep{2003MNRAS.344.1000B} assuming a
Chabrier IMF \citep{ChabrierIMF}. 
As done for $M_I$, we compare the stellar mass function of the $m_\text{F814W}\le23.5$ sample with that of the $m_\text{F814W}\le25.2$ sample. 
The sample with $\log(M/M_\odot) > 10.15$ is $\sim 95$ per cent
complete in the redshift bin $0.75 \le z< 0.85$ and has 8953
galaxies in total across all redshifts.
Thus, we construct a third volume-limited sample \s$3$ by imposing the
stellar mass cut mentioned above. 

The numbers of galaxies in redshift slices
are tabulated in Table~\ref{table:GalaxyCounts} for all three ways
discussed in this section of obtaining a volume-limited sample. The
stellar-mass limited sample is the smallest, most likely because when
converting from flux to stellar mass, the stellar mass-to-light ratios
vary strongly with galaxy type, so red galaxies with high $M_*/L$
simply have too low a flux compared to the blue galaxies at the same
$M_*$, and are not observed.

\begin{table} 
\centering
\begin{tabular}{|r|c|c|c|c|}
 \hline
 Redshift & Environment & \s$1$ & \s$2$ & \s$3$ \\
 \hline
 0.3-0.4 & Overdense & 1726 & 2505 & 1260 \\
 0.4-0.475 & Neutral & 988 & 1312 & 708 \\
 0.475-0.55 & Neutral & 1410 & 1793 & 904 \\
 0.55-0.65 & Underdense & 1797 & 2193 & 1183 \\
 \new{0.65-0.7} & \new{Overdense} & \new{2096} & \new{2321} & \new{1354} \\
 \new{0.7-0.75} & \new{Overdense} & \new{1963} & \new{2155} & \new{1239} \\
 0.75-0.8 & Underdense & 1159 & 1196 & 675 \\
 0.8-0.85 & Overdense & 2428 & 2428 & 1630 \\
 \hline
\end{tabular}
\caption{List of different redshift bins, their environmental
  classification and the number of galaxies per redshift bin for
  volume-limited samples constructed in three different ways: using a
  hard luminosity cut (\s$1$), using a redshift-dependent luminosity cut (\s$2$) and using stellar-mass cuts (\s$3$).}
\label{table:GalaxyCounts}
\end{table}

There is one subtlety in our method used for estimating
completeness. We have used the full $m_\text{F814W}\le23.5$ sample
for identifying
overdensities and for the completeness calculations that motivated our
definitions of volume-limited samples.  However, everywhere else in the paper, we consider only those galaxies for which there are postage stamp
images used to create weak lensing simulations, in part because this is the
sample for which fits to \sersic profiles were carried out, which is a
requirement for our morphology analysis.  
12 per cent of the galaxies that pass our cuts do not have an
associated postage stamp image.
Postage stamps may not exist because, given the size of the galaxy,
the size of the postage stamp we want to draw around it (including
some blank space) intersects the edge of the CCD.
If all galaxies were the same size, this would be a purely random effect, but in fact bigger galaxies are more likely to get excluded by this cut. 
It is commonly the case that  galaxies that are nearby and
intrinsically very bright do not have postage stamps associated with
them, an effect that is dominant at lower redshifts (and is part of
our reason for excluding $z<0.3$). 
Our completeness calculation is done at high redshifts, and thus we
believe that our conclusions are not affected by this bias.  

\begin{figure}
 \centering
  \includegraphics[width=\columnwidth]{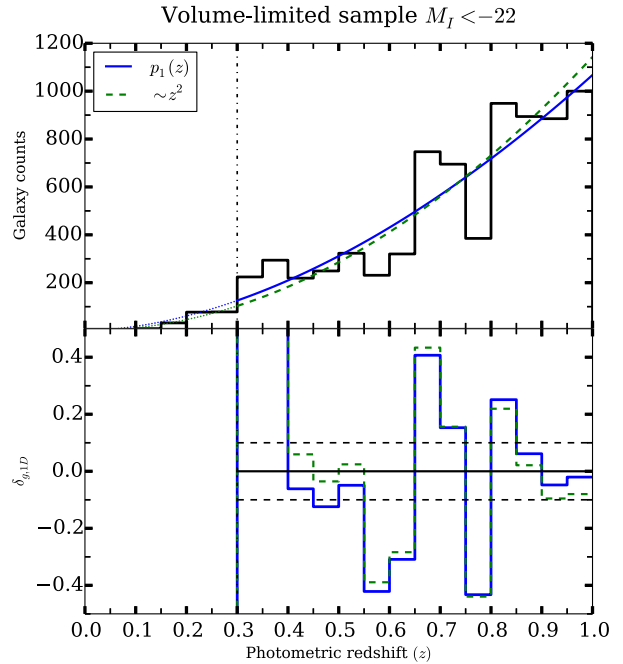}
  \caption{
Upper panel: Redshift distribution of volume-limited ($M_I <  -22$)
sample with photometric redshift bins that are 0.05 wide. Two
analytical functions with best fit parameters are plotted over it, as
discussed in the text. 
           Lower panel: Plot of $(1+\delta_{g,\text{1D}}) =
           N/N_{\text{mod}}$ with each functional form as the model
           for each redshift bin. } 
  \label{fig:redshift_vollimited}
\end{figure}

The functional forms for the (flux-limited) redshift distribution that
we used in Sec.~\ref{sub:overdensities} are not well-motivated for a
volume-limited sample. If we fit them to the 
redshift distribution of the volume-limited sample ($M_I < -22$) that doesn't take into account the evolution of the luminosity function,
then due to the absence of an exponential tail in the histogram, the
parameters that set the scale for the redshift ($z_1$ and $z_2$)
become very large.  As a result, both  $p_1(z)$ and $p_2(z)$ defined
in Eqs.~\eqref{E:pz1} and~\eqref{E:pz2} essentially become the same power law.
The best-fitting exponent is remarkably close to $2$ ($a=2.78\pm
0.28$, giving an exponent $1.78\pm 0.28$), suggesting that the
comoving number density of galaxies is constant with redshift as we
would expect for a volume-limited sample in a redshift range for which
evolution is negligible. 
Fig.~\ref{fig:redshift_vollimited} shows that the values of $\delta_{g,\text{1D}}$ for the $z=0.40-0.55$ bin
increase and are within the $[-0.1, 0.1]$ range that we have defined
as neutral. This is the reason that in Sec.~\ref{sub:overdensities} we classified them as neutral as opposed to underdense.
We will see in Sec.~\ref{S:results} that they are more similar to overdense regions as opposed to underdense regions.
The other redshift slices seem to exhibit a consistent behavior in
Fig.~\ref{fig:redshift_vollimited} and Fig.~\ref{fig:redshift_fluxlimited}. 

\subsection{Describing galaxy morphology and shape}
\label{sub:axisratio}

\begin{figure*}
 \centering
 \includegraphics[width=2.2\columnwidth]{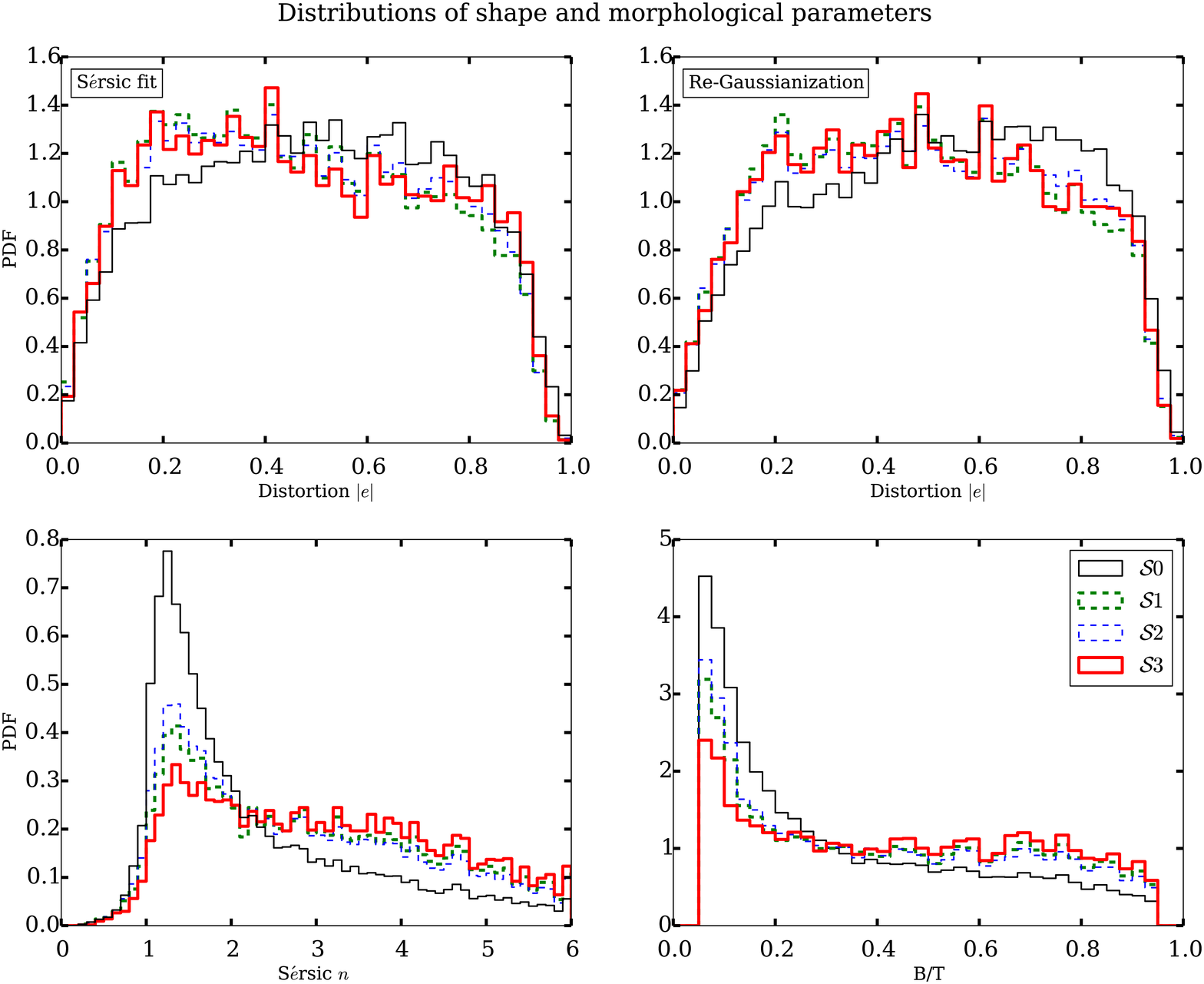}
 \caption{The distributions of the morphological parameters of interest, namely the distortion (top left, from \sersic fits, and top right, from re-Gaussianization), \sersic index (bottom left) and bulge-to-total ratio (bottom right) are presented. 
          The shapes of these distributions depend on the way the volume-limiting process is carried out. \s$0$ refers to the flux-limited ($m_\text{F814}\le23.5$) sample
          and \s$1$, \s$2$ and \s$3$ refer to the volume-limited
          samples, as discussed in
          Sec.~\ref{sub:volumelimiting}. 
  }
  \label{fig:sample_histograms}
\end{figure*}

We choose simple and well-motivated ways to parametrize galaxy shapes
and morphology based on existing methods in the literature.  These
methods have the advantage of being stable and well-defined in nearly
all cases.  However, for highly irregular galaxies the meaning of the
structural parameters that we derive is not entirely clear.  In all
cases, our methods account for the effect of the {\em HST} PSF.

One method to estimate the galaxy ellipticities and other
morphological parameters is to fit parametric models convolved with
the PSF to the observed galaxy light profile.  We use the fits from
\cite{great3}, which used the methods and
software from \cite{Claire_Fits} to fit the images to the following profiles:
\begin{enumerate}
 \item A \sersic profile given by the expression 
       \begin{equation} 
    I_S(x,y) = I_{1/2}\exp{\left[ -k(R(x,y)/R_{\text{eff}})^{1/n} -1 \right]},
       \end{equation}
where \begin{align*} R^2(x,y) = & ((x-x_0)\cos\Phi+(y-y_0)\sin\Phi)^2
  \\ & + ((y-y_0)\cos\Phi-(x-x_0)\sin\Phi)^2/q^2, \end{align*}
$R_{\text{eff}}$ is the half-light radius of the profile defined along
the major axis, $I_{1/2}$ is the surface brightness at $R=R_{\text{eff}}$, $(x_0,y_0)$ is the centroid of the image,
$\Phi$ is the position angle, $n$ is the \sersic index, $k$ is a
$n$-dependent normalization factor required to ensure that half the
light is enclosed within the half-light radius, and $q$ is the axis
ratio of the elliptical isophotes.
Thus, the \sersic profile has 7 free parameters.
       \item A sum of two \sersic component fits: a de Vaucouleurs
         bulge $(n=4)$ plus an exponential disc profile $(n=1)$.  In this
         case, there are 10 free parameters, because the \sersic\
         indices are fixed, and the two components are constrained to
         have the same centroid.
\end{enumerate}
More details about the fitting algorithm can be found
in \cite{Claire_Fits}.

To quantify galaxy morphology and shape, we will use several
quantities from the above fits.  First, from the single \sersic
profile fits, we use the \sersic index and the axis ratio.  The axis
ratio can also be used to derive a distortion,
\begin{equation}\label{eq:distortion}
e = \frac{1-q^2}{1+q^2}
\end{equation}
or an ellipticity,
\begin{equation}\label{eq:ellipticity}
\varepsilon = \frac{1-q}{1+q}.
\end{equation}

As an alternative morphological indicator (instead of \sersic index)
we use a bulge-to-total ratio derived from the double \sersic
profile fits.  This ratio is defined in terms of bulge and disk fluxes as
\begin{equation}
\frac{B}{T} = \frac{F_\text{bulge}}{F_\text{bulge}+F_\text{disk}}.
\end{equation}
The bulge-to-total flux ratio can be used as a proxy for
  colour gradients, since the bulge and disk will tend to have
  different spectral energy distributions, and hence any galaxy with
  $B/T\ne 0$ or $1$ will have some level of colour gradients.  If two
  galaxy samples have different values for the typical $B/T$, this is
  likely to indicate not only differences in morphology, but also in
  the level of colour gradients.
 
We also consider an alternative method for estimating the galaxy
ellipticity or distortion.  This method is based on using the observed weighted
moments of the galaxy and PSF, and correcting those of the galaxy for
those of the PSF.  This PSF correction scheme is the
re-Gaussianization method described in section~2.4 of \cite{HS03} 
as implemented in the {\sc GalSim} software package
(with implementation details described in \citealt{2014arXiv1407.7676R}).
This method models the true PSF $g({\bf x})$ as a Gaussian $G({\bf x})$ and the residual $\epsilon({\bf x}) = g({\bf x}) - G({\bf x})$ is assumed to be small. Thus, the Gaussian-convolved
intrinsic image, $f$, can be modeled as $I' = G\otimes f = I - \epsilon \otimes f$, where $I$ is the observed image. The crucial idea here is that, when $\epsilon$ is small, we get a reasonably accurate
estimate of $I'$ even if we use an approximate form for $f$. The form
assumed for $f$ is that of a Gaussian with covariance $M_f^{(0)} = M_{(I)}
- M_{(g)}$, where $M_{(I)}$ and $M_{(g)}$ are the elliptical Gaussian-weighted adaptive
covariances of the measured object and PSF respectively, described in
section~2.1 of \cite{HS03} and \cite{BJ02}. We refer to the
re-Gaussianization estimates of the PSF-corrected distortion as
``moments-based shape estimates''.  The value in including them in
this analysis is that they have quite different radial weighting from
the \sersic profile fits, with the outer regions being quite
downweighted when calculating adaptive moments.  Thus, if ellipticity
gradients are important, we could get different results using these
two shape estimators.

Fig.~\ref{fig:sample_histograms} shows the distribution of these morphological and shape parameters for the flux-limited sample \s$0$ and the three
volume-limited samples - \s$1$, \s$2$ and \s$3$. In addition to the
basic value in characterizing these distributions for our sample
overall, it is also useful to understand how the samples change when
we vary our method of volume-limiting the sample.
For instance, galaxies with lower \sersic indices are preferentially
selected in \s$2$ compared to \s$1$ and \s$3$. Similarly, galaxies
with low ellipticity/distortion values 
are rejected in generating \s$2$ (despite the lack of any explicit
cut on shapes) while they are retained in
\s$1$ and \s$3$.  A simple explanation is that the cuts in \s$2$ are
preferentially removing early-type populations, which have higher
\sersic indices and lower ellipticities.  For example, if the
luminosity evolution that was adopted is too strong particularly for
early type populations, that could give rise to the effect shown in
Fig.~\ref{fig:sample_histograms}.

\section{Results}
\label{S:results}

Having identified the overdense and underdense regions in a
volume-limited sample (Secs.~\ref{sub:overdensities}
and~\ref{sub:volumelimiting}), we will now see whether the morphological
parameters of the galaxies described in Sec.~\ref{sub:axisratio}
depend noticably on the environment of the redshift slice in which
they reside. Note that for true 3D overdensities there is already
substantial evidence in the literature that we should see variation of
properties with the environment.  Our test is necessary to see
whether such morphology-density correlations are evident in the kind
of 1D redshift slices that would be used for constructing weak lensing
simulations, or whether our use of an area as large as the size of COSMOS will
wash out these trends (which would be good news for weak lensing
simulations based on that dataset).

As described in Sec.~\ref{sub:volumelimiting}, we have three different ways of volume-limiting our sample:
\begin{enumerate}
 \item no redshift evolution of luminosity cut $(\s1)$,
 \item using $B$-band luminosity evolution applied to the $I$-band
   luminosities $(\s2)$, and
 \item impose stellar mass cuts instead of luminosity $(\s3)$.
\end{enumerate}

We will present our results in all three cases to check for their
robustness to how the sample is selected. 

\subsection{Axis ratios} 

We can test the influence of environment on the galaxy shapes by
comparing the distributions of the axis ratios for the overdense and
underdense redshift slices, or by encapsulating that distribution as a
single number, the RMS (root mean squared) ellipticity or distortion. By
volume-limiting the sample, we have avoided issues
wherein the flux limit leads to artificial changes in the sample as a
function of redshift.  We will also carry out tests to differentiate
between environmental effects versus evolution of the population with
redshift (at fixed mass).

\subsubsection{Comparing distributions}
\label{subsubsec:compdist}
\begin{figure*}
 \centering
 \includegraphics[width=2.2\columnwidth]{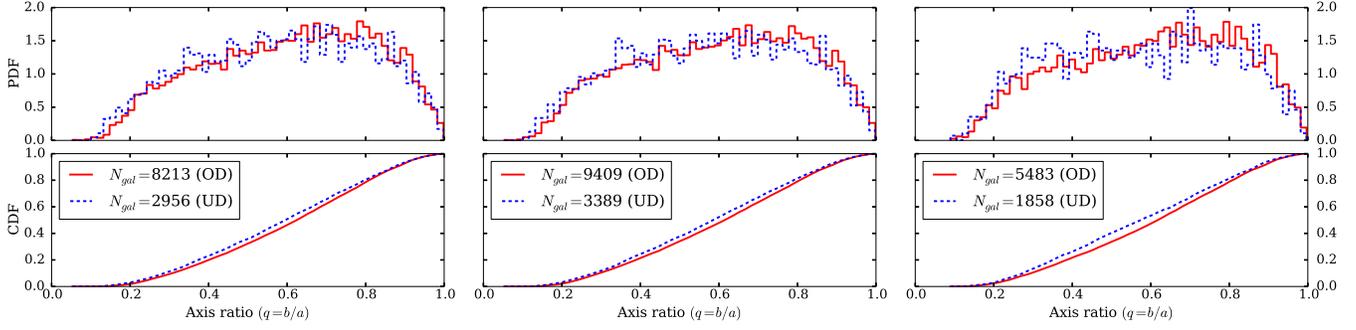}
 \caption{The distributions of axis ratios of galaxies in \emph{all}
   overdense (OD) and \emph{all} underdense (UD) regions in the case
   of the luminosity-selected sample \s1 (left), luminosity-selected
   sample with $B$-band evolution taken into account \s2 (center), and
   thestellar-mass-selected sample \s3 (right). The upper panels show
   the histograms, 
 and the bottom panels show the cumulative distribution functions
 (CDF). The $p$-values computed using these CDFs are shown in Table. \ref{table:pvalues_all}.}
 \label{fig:axisratio_all}
\end{figure*}

\begin{figure*}
 \centering
 \includegraphics[width=2.2\columnwidth]{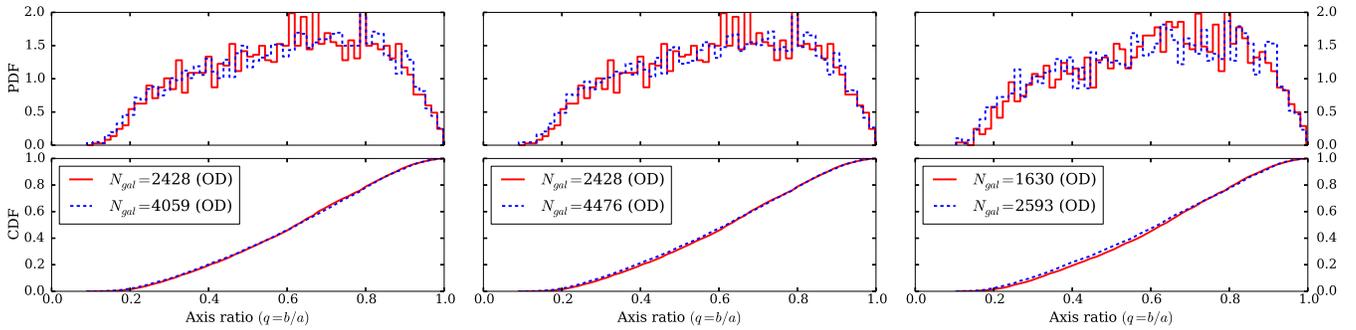}
 \caption{Galaxy axis ratio
   distributions in two overdense redshift bins, $z=0.65-0.75$ (blue, dotted) and
   $z=0.80-0.85$ (red, solid), to check for consistency in the case that the
   environment is the same even if the redshift differs. The
   $p$-values from the KS and AD tests are given in
   Table.~\ref{table:pvalues_all}.}
 \label{fig:axisratio_similar}
\end{figure*}

\begin{figure*}
 \centering
 \includegraphics[width=2.2\columnwidth]{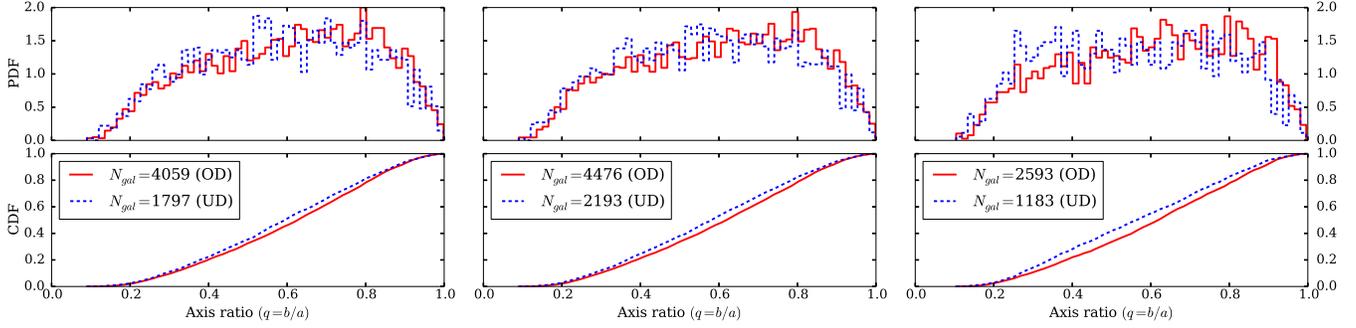}
 \caption{Galaxy axis ratio distributions in a single underdense redshift
   slice, $z=0.55-0.65$ (blue, dotted), and a single overdense redshift slice,
   $z=0.65-0.75$ (red, solid). The $p$-values from the
   KS and AD tests are given in Table.~\ref{table:pvalues_all}. }
 \label{fig:axisratio_contrasting}
\end{figure*}

\begin{table}
 \centering
 \begin{tabular}[\columnwidth]{ | l | c | c | c | }
  \hline
  Redshift bins & \s$1$ & \s$2$ & \s$3$ \\
  \hline
  All overdense vs.\  & \scinot{1.1}{-4} & \scinot{2.6}{-5} & \scinot{1.9}{-6} \\
  All underdense     & \scinot{1}{-5} & $<\scinot{1}{-5}$ & $<\scinot{1}{-5}$ \\ \hline 
  $[0.65,0.75]$ (OD) vs.$\!\!\!$ & 0.61 & 0.43 & 0.23 \\
  $[0.80,0.85]$ (OD) & 0.49 & 0.24 & 0.13 \\ \hline
  $[0.65,0.75]$ (OD) vs.$\!\!\!$ & \scinot{5.8}{-4} & \scinot{1.5}{-5} & \scinot{3.5}{-6} \\
  $[0.55,0.65]$ (UD) & \scinot{9.8}{-4} & $<\scinot{1}{-5}$ & $<\scinot{1}{-5}$ \\ \hline
 \end{tabular}
 \caption{$p$-values from the Kolmogorov-Smirnov (top) and
   Anderson-Darling (bottom) tests obtained by comparing the
   distributions of axis ratios for three cases: \emph{all} overdense
   (OD) vs.\ \emph{all} underdense (UD), two overdense bins that are
   not very separated in redshift, and a pair of adjacent overdense
   and underdense bins. \s$1$, \s$2$, \s$3$ refer to the three
   different types of volume-limited samples.
          The Anderson-Darling $p$-values are computed only up to 5
          decimal places, so values that were given as zero are
          denoted $<\scinot{1}{-5}$.
          }
 \label{table:pvalues_all}
\end{table}

\begin{table}
 \centering
 \begin{tabular}[\columnwidth]{ | l | c | c | c | }
  \hline
  Redshift bins & \s$1$ & \s$2$ & \s$3$ \\
  \hline
  All overdense vs.\ & \scinot{5.6}{-4} & \scinot{1.0}{-4} & \scinot{3.3}{-6} \\
  All underdense    & \scinot{3}{-5} & \scinot{1}{-5} & $<\scinot{1}{-5}$ \\ \hline 
  $[0.65,0.75]$ (OD) vs.$\!\!\!$ & 0.96 & 0.75 & 0.54 \\
  $[0.80,0.85]$ (OD) & 0.52 & 0.34 & 0.23 \\ \hline
  $[0.65,0.75]$ (OD) vs.$\!\!\!$ & \scinot{6.0}{-3} & \scinot{2.5}{-4} & \scinot{2.4}{-4} \\
  $[0.55,0.65]$ (UD) & \scinot{1.2}{-2} & \scinot{2.5}{-4} & \scinot{5}{-5} \\ \hline
 \end{tabular}
 \caption{$p$-values from the Kolmogorov-Smirnov (top) and
   Anderson-Darling (bottom) obtained by comparing the second
   moments-based distortion for the same three cases as in
   Table~\ref{table:pvalues_all}. 
           The Anderson-Darling $p$-values are computed only up to 5 decimal places. }
 \label{table:pvalues_momentbased_all}
\end{table}

We begin by comparing the entire axis ratio distributions $p(q)$
between pairs of redshift slices.  Unless otherwise mentioned, the
axis ratios refer to the values obtained using the method of
\cite{Claire_Fits} to fit single \sersic profiles to each galaxy image.
To compare the distributions and
make statistical statements about their consistency, we
use two statistical tests, the Kolmogorov-Smirnov (KS) test and
Anderson-Darling (AD) test, the latter of which is carried out using
the \texttt{adk} package in \texttt{R}.

We first compare the distribution of galaxy axis ratios in \emph{all}
overdense bins against that for \emph{all} underdense bins in
Fig.~\ref{fig:axisratio_all}, with different panels showing the
comparison for \s1, \s2, and \s3. 
The cumulative distribution functions are also shown, since they form
the basis for our statistical statements about consistency using the
KS and AD tests.  The results of these tests are shown in the first
two rows in Table~\ref{table:pvalues_all}.  For all three ways of
volume-limiting the sample, the $p$-values from both the KS and AD
tests are well below 0.05 (a maximum of $1.1\times 10^{-4}$, but often
smaller than that).  We can therefore reject the null hypothesis that
the overdense and underdense regions have the same underlying axis
ratio distributions at high significance.

One might imagine that the disagreement between the distributions is,
at least partly, due to the fact that the overdense and underdense
sample have different redshift distributions and there could be some
evolution of ellipticity/distortion distributions with redshift. 
To show that this redshift evolution effect is subdominant to
environmental effects, we will compare distributions between pairs of
two overdense (or pairs of underdense) redshift slices, where we
expect to find similarity even if the redshifts are different if the
environmental effects dominate.  We will also compare between
overdense and underdense regions that are selected to be nearby in
redshift, so that any redshift evolution effects should be minimal. 
Figures~\ref{fig:axisratio_similar}
shows that the axis ratio
distributions are indeed consistent when the environments are similar
but the redshifts are different.  Likewise
Fig.~\ref{fig:axisratio_contrasting} shows that for adjacent redshift
slices with different environments, the axis ratio distributions are
inconsistent.  The results of statistical tests for the distributions
in these figures are given in Table~\ref{table:pvalues_all}, and
support our statement that the morphology-density
correlation is the dominant effect when comparing overdense and
underdense redshift slices, with redshift evolution of the
population being negligible. 
Comparing other pairs of redshift
bins leads to similar conclusions. 

Finally, we can check whether these findings are particular to the
axis ratios from the \sersic fits, or whether we reproduce this
finding when we use the shapes from the centrally-weighted
moments-based re-Gaussianization method, which estimates a distortion
(Eq.~\ref{eq:distortion}) for each galaxy. 
After neglecting a small fraction ($<0.01$ per cent) of galaxies for
which the method does not converge, 
we carry out the same statistical tests from
Table~\ref{table:pvalues_all}, but using the moments-based shape
estimates.  The results of the KS and AD tests are tabulated in Table~\ref{table:pvalues_momentbased_all}.
We see that all of our findings with the \sersic fit-based shapes
carry over to shapes from a centrally-weighted moments-based shape
estimate.

\subsubsection{RMS distortions}
\begin{figure}
 \centering
 \includegraphics[width=1.0\columnwidth]{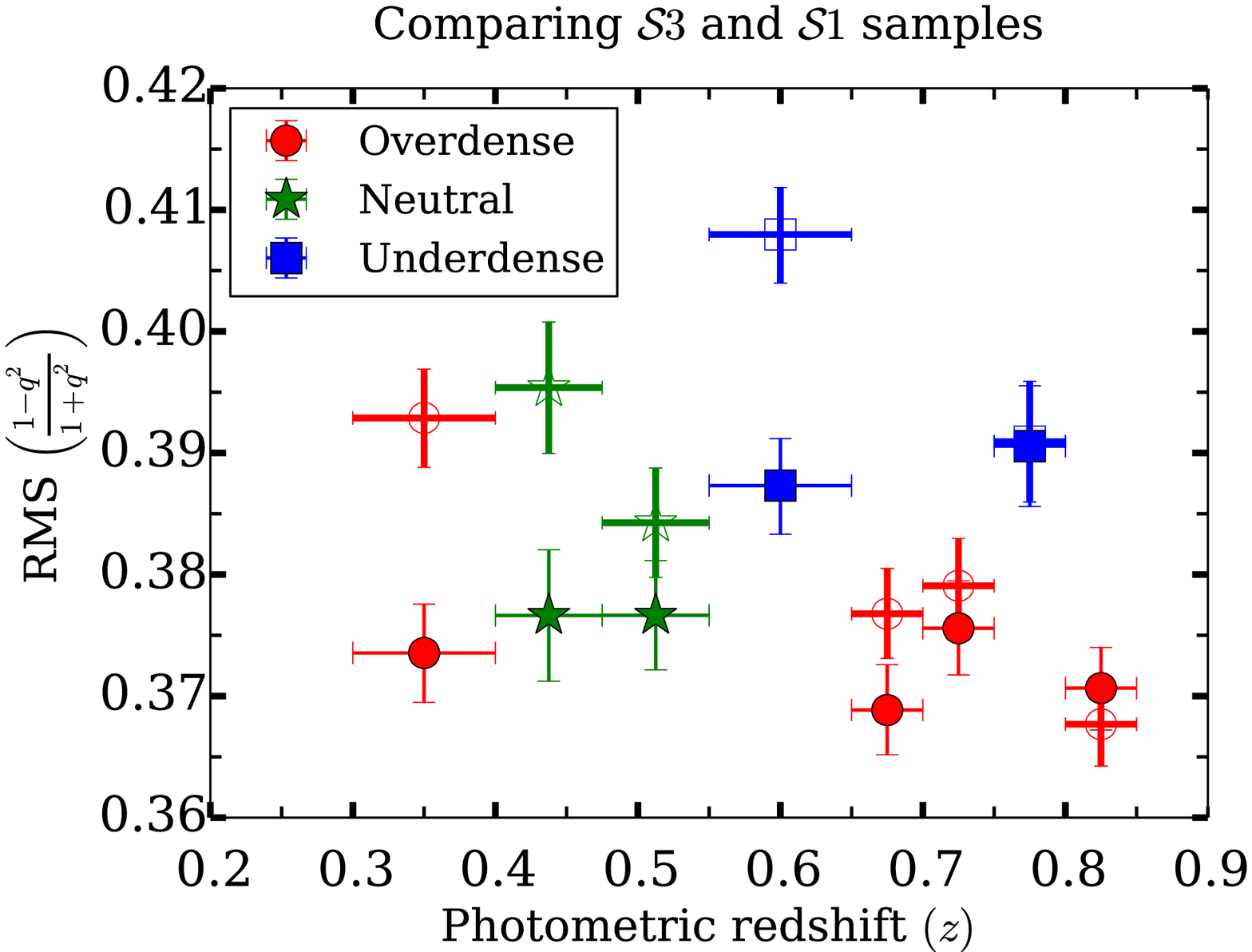} \\
 \includegraphics[width=1.0\columnwidth]{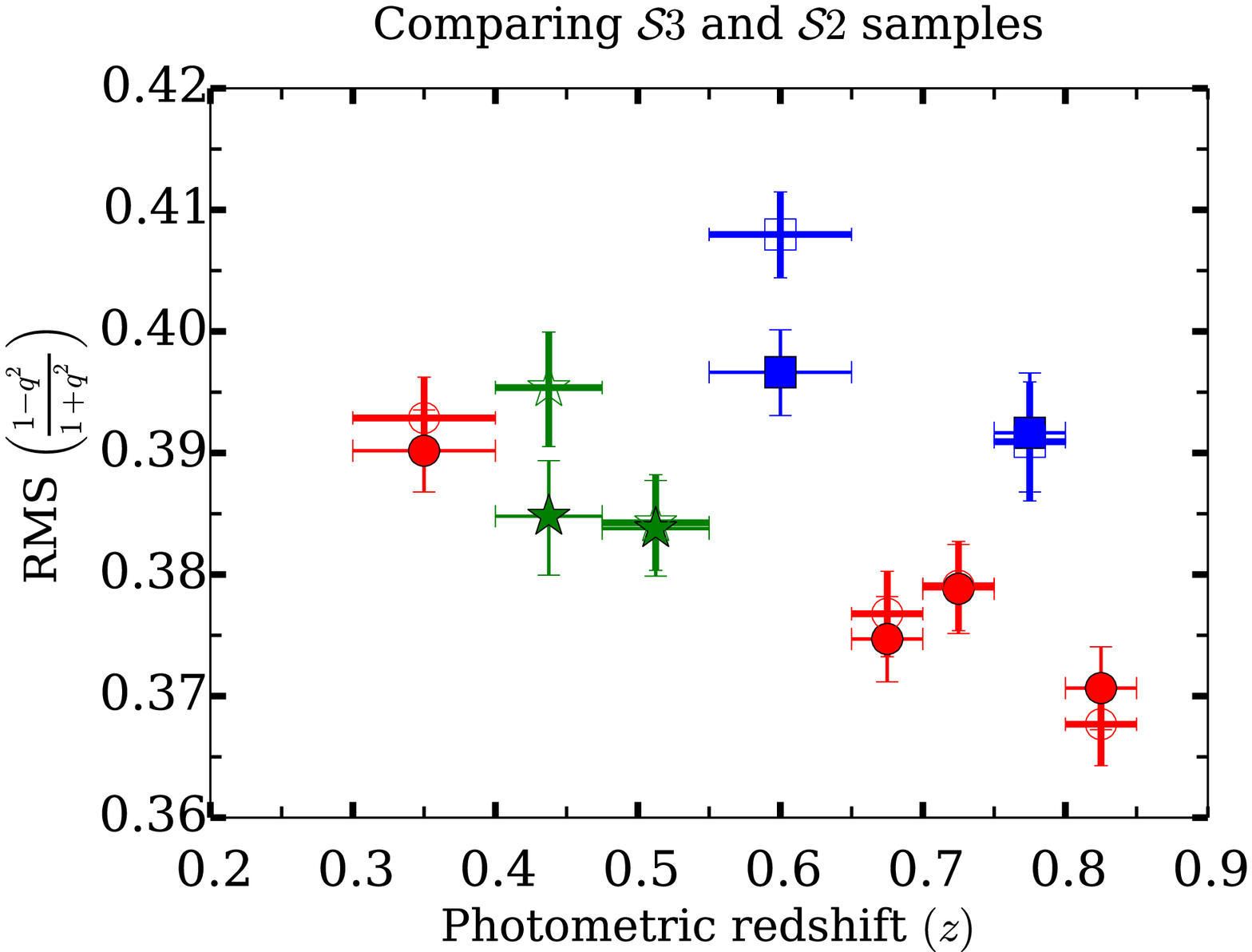} \\
 \caption{RMS distortions as a function of redshift. The horizontal errorbars indicate
          the width of the redshift bin, while the vertical ones
          are $1\sigma$ errorbars obtained by
          bootstrapping. The colours and shapes of the points indicate their
          environmental classification, as shown in the legend. Points with open centers and thick errorbars correspond to the stellar-mass selected sample \s$3$
          and points with filled centers and thin errorbars correspond to the luminosity-selected samples \s$1$ and \s$2$.             
          }
 \label{fig:rms_ellip}
\end{figure}

\begin{figure}
 \includegraphics[width=1.0\columnwidth]{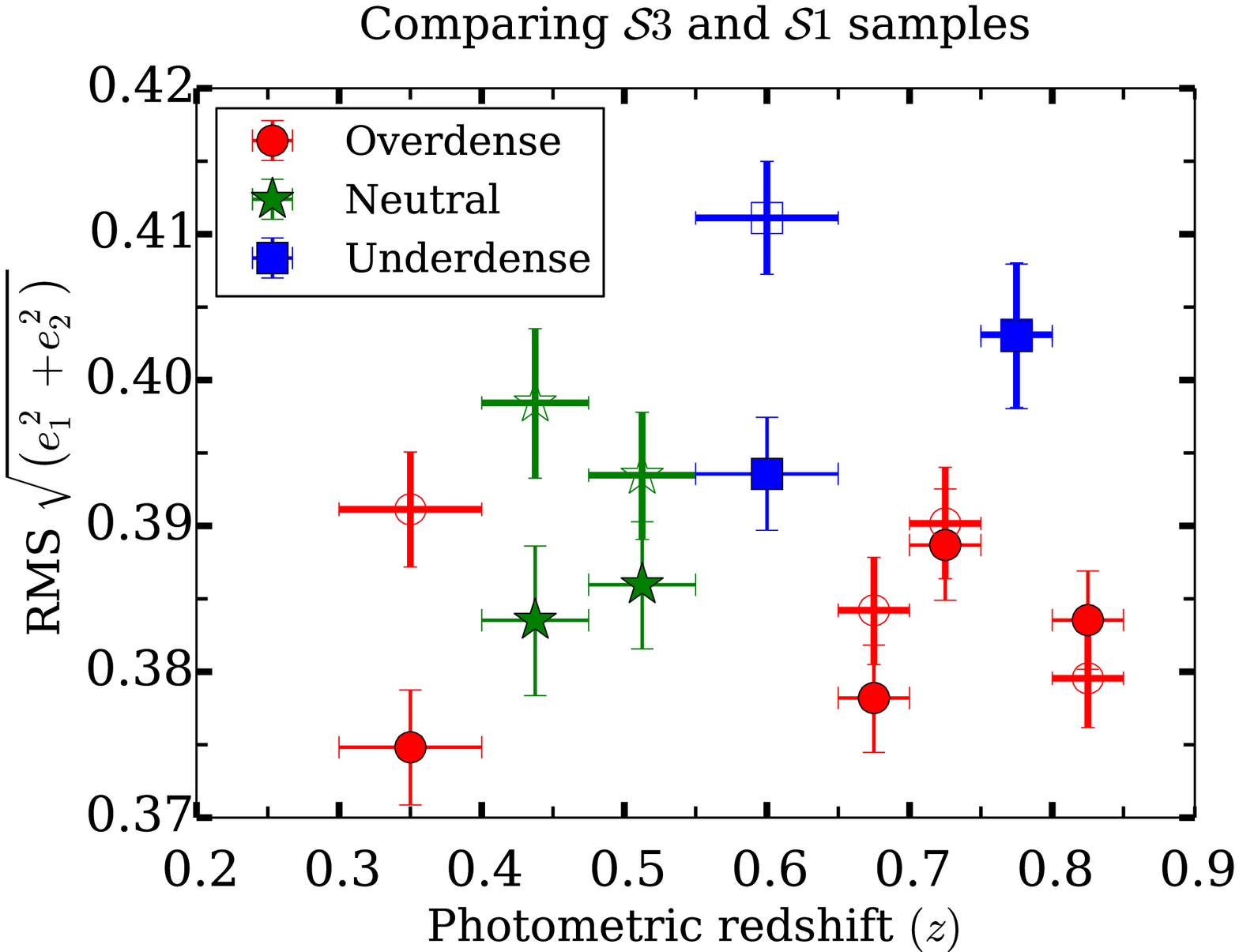} \
 \includegraphics[width=1.0\columnwidth]{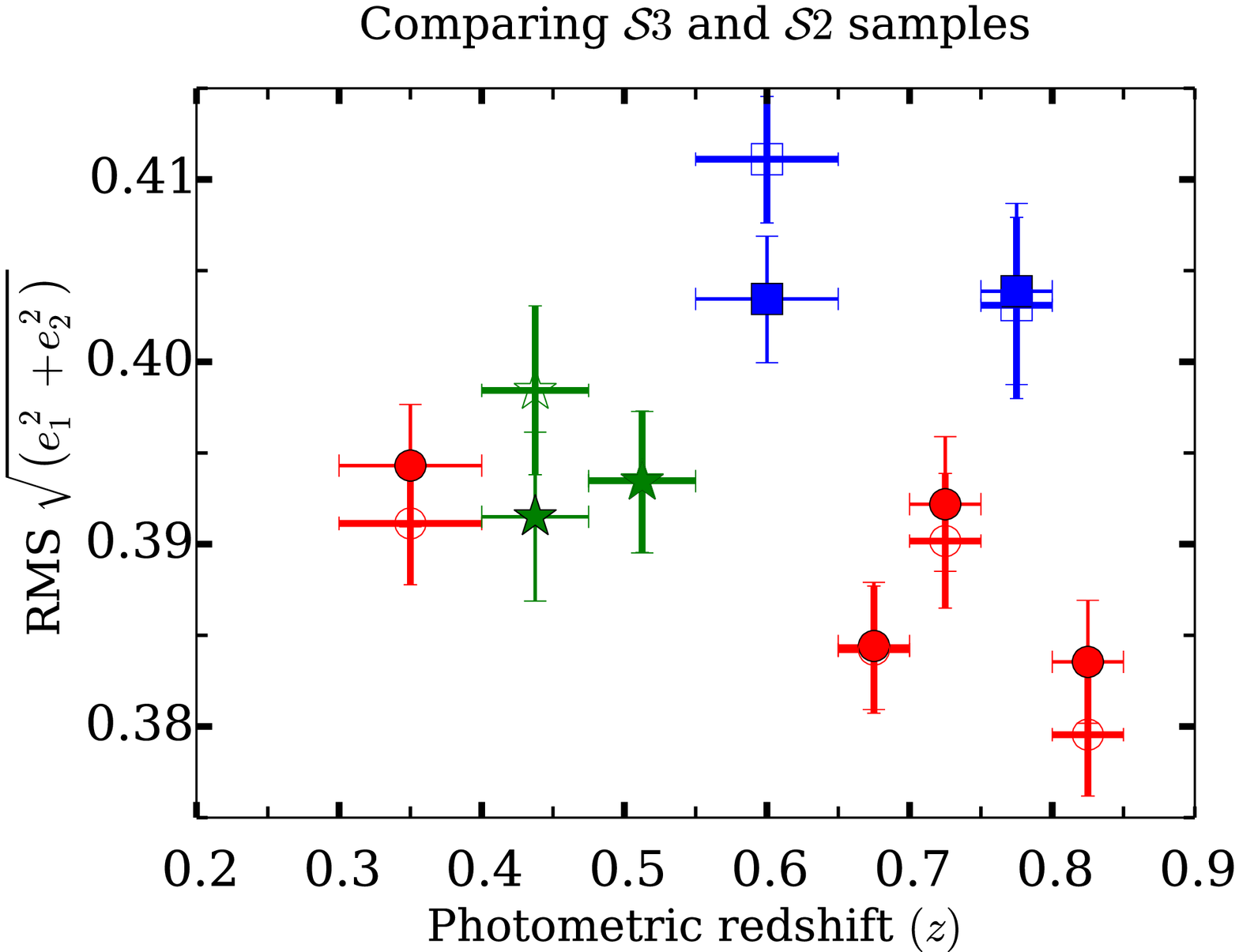} \\
 \caption{RMS distortions as a function of redshift, with points
   defined in a similar way as in Fig.~\ref{fig:rms_ellip}.  In this
   case, the distortions from moments-based shape estimates, rather
   than from the \sersic fits. }
 \label{fig:rms_ellip_momentbased}         
\end{figure}

We can also carry out tests on a single statistic of the galaxy shape
distribution in each redshift slice, like the RMS distortion.  While
tests of a single quantity may seem less powerful than tests that use
the entire shape distributions, the advantage is that instead of
picking out pairs of redshift slices for our tests, we can easily compute our
statistic of interest for every single redshift slice, and look for
trends with both redshift and environment.

For the luminosity-selected samples (\s$1$, \s$2$), the
RMS distortions (Eq.~\ref{eq:distortion}) of galaxies in each redshift bin are shown
in Fig.~\ref{fig:rms_ellip}. In each case, the RMS distortions from the mass-selected sample (\s$3$) are also plotted.
When the $B$-band luminosity evolution is taken into account in selecting the sample, a systematic increase in the distortion values at
lower redshifts can be observed (bottom panel).The stellar-mass selected sample exhibits a similar trend.

For figures up to Fig.~\ref{fig:median_dvcbtt}, the colours of the points were selected to easily differentiate between galaxies in overdense, neutral, and underdense redshift slices.
Points with unfilled centers and thicker errorbars correspond to the \s$3$ sample and points with filled centers and thinner errorbars correspond to the luminosity-selected samples (\s$1$ or \s$2$).

As shown in Fig.~\ref{fig:rms_ellip}, the underdense regions have higher values for RMS distortions when compared to the
overdense regions. The difference between the underdense and overdense
regions for $z>0.5$ is significantly larger than any redshift
evolution across the $z>0.5$ range.  
Our conclusions are very similar if we use the RMS ellipticity from Eq.~\eqref{eq:ellipticity} instead
of the distortions.

The sign of the dependence on the local environment is reasonable when
compared with previous work on the morphology-density
relation \citep[see, e.g.][]{2010ApJ...714.1779V}. Overdense regions typically contain many old, elliptical
galaxies which are close to round (large axis ratio and low RMS ellipticity/distortion). 
In contrast, the underdense regions typically contain a larger
population of younger, disk galaxies, which have lower axis ratios and
higher RMS ellipticity/distortion.

From Figs.~\ref{fig:redshift_fluxlimited} and~\ref{fig:redshift_vollimited},
the $0.4\le z < 0.55$ redshift range shows signs of being marginally
underdense, but has low RMS ellipticity that agrees with the rest of the overdense regions.

Next, we show an analogous plot of RMS distortions for all three
volume-limited sample using the moments-based shape estimates in
Fig.~\ref{fig:rms_ellip_momentbased}. The conclusions are quite
similar to those using the shapes from the \sersic profile fits, with
the underdense regions standing out in having larger RMS distortions 
than the other redshift slices, an effect that is substantially larger
than any average redshift evolution of the RMS distortions.

However, the statistical significance of
trends in this section using a single statistic of the shape
distribution (the RMS) is less than the significance of the trends
seen using the entire axis ratio distributions in Sec.~\ref{subsubsec:compdist}.

Finally, we comment on whether the trends we
  have found could be caused by measurement noise rather than
  variations in the intrinsic shape distributions.  Measurement error
  tends to cause an increase in the RMS ellipticity due to the
  broadening of the measured ellipticity distributions.  However, this
is one of the reasons we restricted the sample to a magnitude of
23.5.  In this case, the $S/N$ of the flux measured in the galaxy
images is typically $\ge 50$.  If we consider the worst case, i.e.,
assume all galaxies have $S/N=50$ and use the Gaussian approximation
from \cite{BJ02}, the expected measurement error on the distortion is
$0.08$.  If we add this in quadrature with an RMS of $\sim 0.33$, then
the new RMS becomes $0.34$.  However, we see variations in the RMS
distortion in different environments that are as large as $0.05$, or
five times as large as this worst-case scenario due to measurement
noise.   Moreover, if measurement error were a significant issue, we
would expect it to affect the comparison of (for example) low- and
high-redshift samples in overdense regions.  In a volume-limited
sample, those will have different flux distributions and therefore
different SNR distributions.  However, there is little trend in the
RMS distortion with redshift for overdense regions, which suggests
empirically that measurement error is not a significant factor in our results.

\subsection{Morphological parameters}

For the morphological parameters that we described in
Sec.~\ref{sub:axisratio}, the \sersic index and bulge-to-total ratio,
we do not compare the distributions directly.  Doing so is relatively difficult 
because both distributions have hard cutoffs that are enforced
in the fitting process (\sersic $n$ in the range $[0.25, 6]$ and
$0.05<B/T<0.95$), as can be seen in Fig.~\ref{fig:sample_histograms}.
The KS statistical test is sensitive to exactly what happens at these hard boundaries in the distributions.
So, instead of using the full distributions, we will study the
dependence of these quantities on environment by computing the median
values in different redshift slices.  Median values are preferred
over the full distributions or even the sample means since the medians
are more robust to what happens at the edges of the distributions.

\begin{figure}
 \includegraphics[width=1.0\columnwidth]{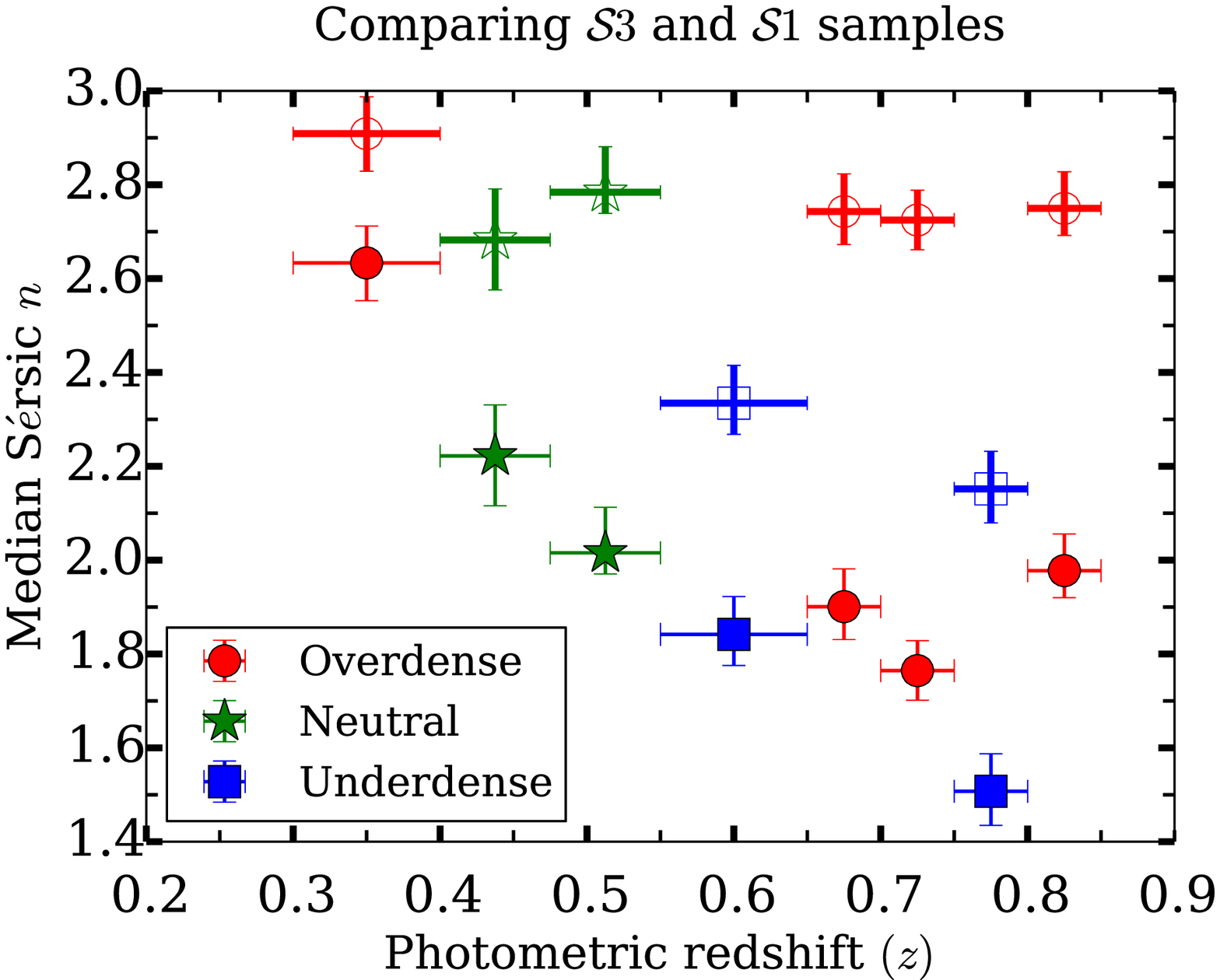}
 \includegraphics[width=1.0\columnwidth]{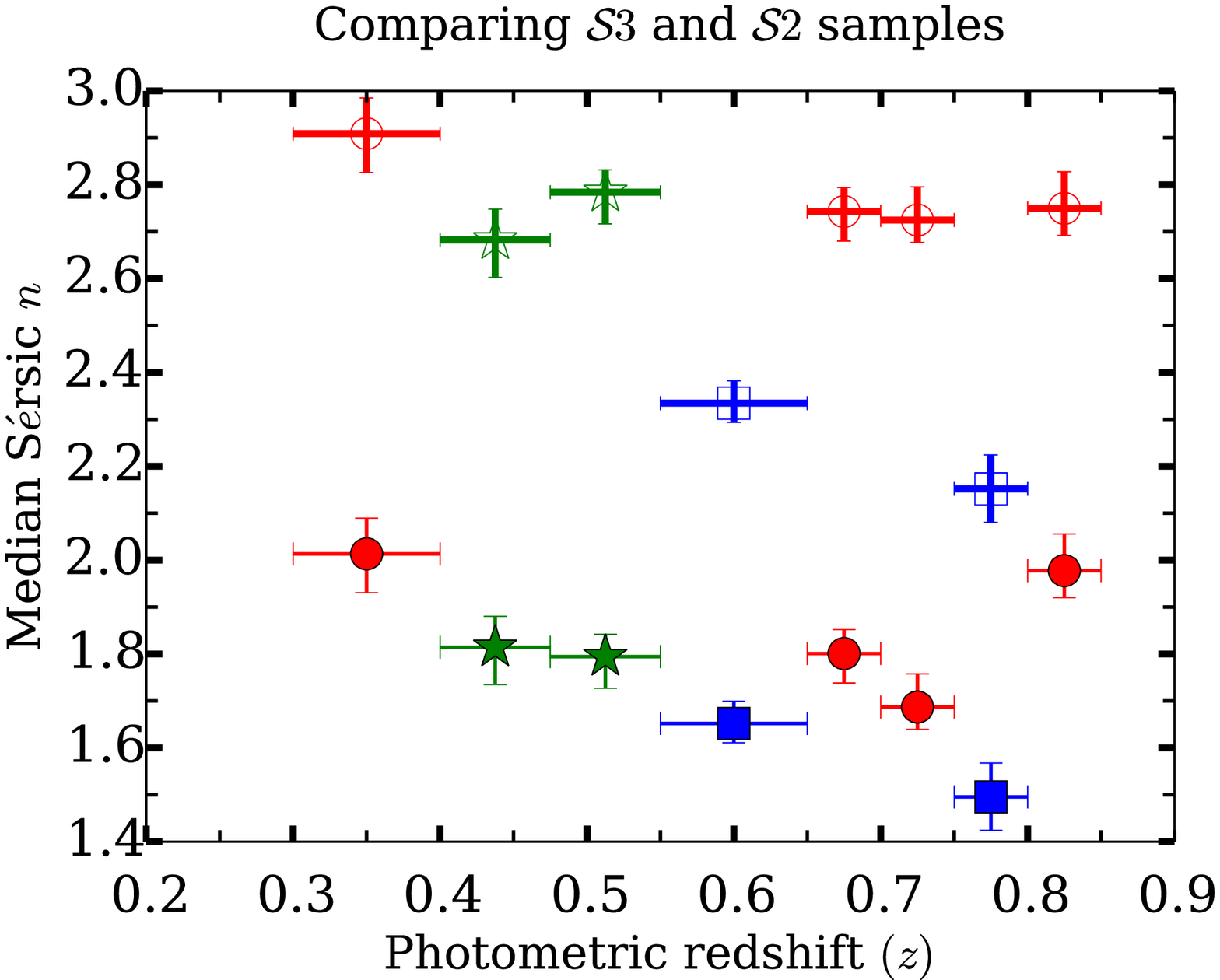}
 \caption{Median values of the \sersic indices for volume-limited
   samples \s$1$ and \s$2$ are plotted (filled centers and thin
   errorbars) in top and bottom panels, respectively, for each
   redshift bin. 
          Median values for the \s$3$ sample are plotted in both the
          panels (open centers and thick errorbars) in both the
          panels.
     }
 \label{fig:median_sersicn}
\end{figure}

Fig.~\ref{fig:median_sersicn} shows the median value of the \sersic
index in each redshift slice, with and without taking into account of
luminosity evolution when volume-limiting the samples (\s1 and
\s2). Both panels also show the results with stellar mass-selected
samples (\s3) for reference.  
When using the stellar mass-selected sample, we observe that the
overdense regions tend to have higher \sersic 
index than the underdense ones, with the redshift evolution being mild
and the results for the underdense regions particularly standing
out. This trend is consistent with our previous explanation for trends
in RMS ellipticities; the underdense regions have more spiral galaxies
and therefore a lower median value of \sersic $n$. 
However, this trend is less evident for the luminosity-selected
samples, where there seems to be some evolution with redshift that
dominates over the environmental effects.

We also note that in Fig.~\ref{fig:median_sersicn}, the median \sersic
indices of the stellar mass-selected samples (\s$3$) are
systematically greater than those of the luminosity selected samples
(\s$1$, \s$2$).  
This is because \s3 is restricted to galaxies with masses above $\log(M/M_\odot) > 10.15$, whereas in \s1 and \s2, the mass distribution of galaxies extends to $\log(M/M_\odot) \sim 9.0$,
with about 44 per cent of the galaxies in \s2 having a stellar mass below the cut for \s3.
It is therefore not surprising that \s3 has a higher median \sersic $n$.  Finally, even for
the stellar mass-selected sample there is some sign of redshift
evolution.  The sign of this evolution is as expected, with lower
\sersic $n$ and $B/T$ for higher redshift samples, which should have a
higher fraction of disk and irregular galaxies and fewer galaxies with
bulge-like morphology.
 
\begin{figure}
  \includegraphics[width=1.0\columnwidth]{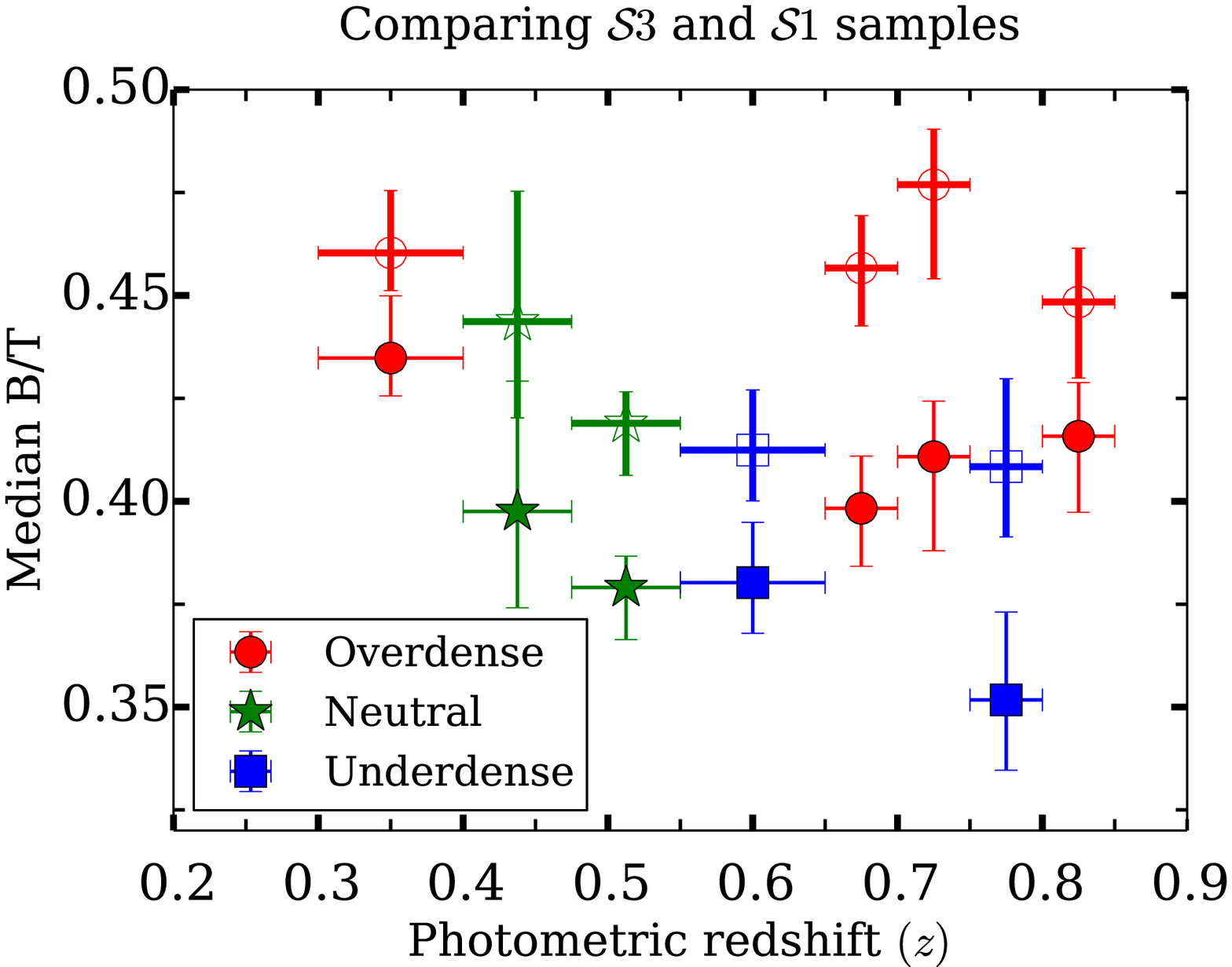} \
 \includegraphics[width=1.0\columnwidth]{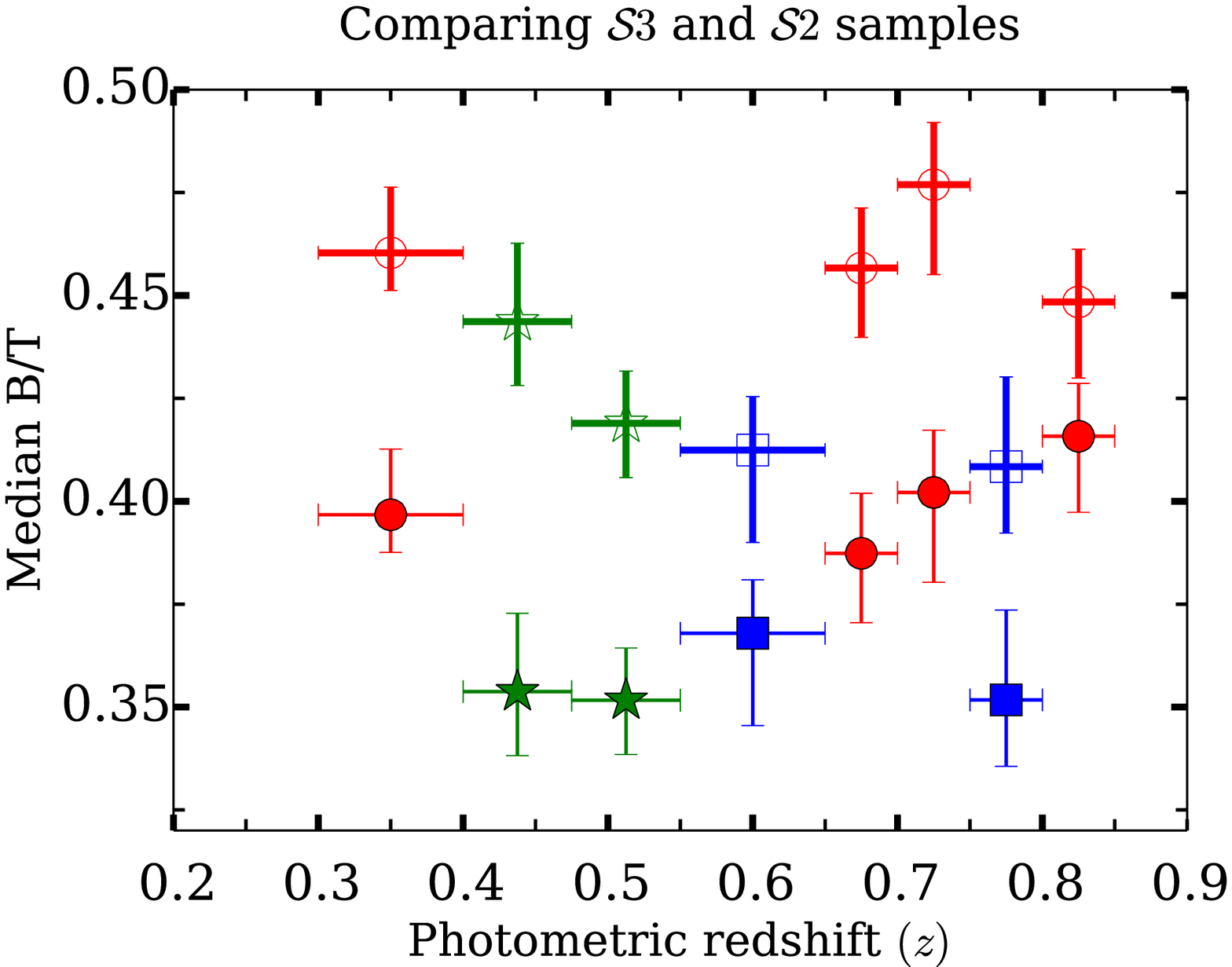}
  \caption{ Median values of the bulge-to-total ratios for volume-limited samples \s$1$ and \s$2$ are plotted (filled centers and thin errorbars) in left and right panels respectively for each redshift bin.
          Median values for the \s$3$ sample are plotted in both the panels (open centers and thick errorbars).
          The horizontal errorbars simply correspond to the binwidth while the vertical ones are $1\sigma$ errorbars obtained by bootstrapping.}
 \label{fig:median_dvcbtt}
\end{figure}

Finally, Fig.~\ref{fig:median_dvcbtt} shows the variation of the median
bulge-to-total ratio with redshift. The results are quite consistent
with those of Fig.~\ref{fig:median_sersicn}. 
Thus, our results in this section suggest that the 
environment can significantly affect the median morphological
parameters of galaxies selected in thin redshift
slices, assuming that the galaxies represent a stellar mass-selected
sample.  The trend is less evident when using luminosity to select the
galaxies.

\new{
One might expect the points corresponding to neutral regions to lie in between the points for overdense and underdense regions. 
In Figs.~\ref{fig:rms_ellip}--\ref{fig:median_dvcbtt}, this does not always appear to be the case. 
It is possible that a redshift bin may have an overdensity in one part of the field and an
underdense region in another. 
In such a scenario, the redshift slice might appear to be `neutral' in our histogram-based method of
finding overdensities, while still having significant large-scale structure affecting the
morphological mix of galaxies that complicates the situation. 
This indeed turns out to be the case
in the `neutral' regions of the COSMOS field.
Using X-ray information in the COSMOS field, \cite{COSMOS_XRAY} have detected several galaxy groups in the
redshift range $z=0.425-0.575$.}
\new{\cite{Kovac_Density10k} report structures that are large and extended (in RA-DEC) within this redshift range. 
It is a challenge to make definitive statements regarding the relationship between environment and
morphology in neutral regions that are composed of mixtures of multiple overdense and underdense regions. 
In addition, the redshift evolution, albeit weak, can make the neutral bins look like overdense or underdense regions.
For instance, the neutral bins look overdense-like in Figs.~\ref{fig:rms_ellip}-\ref{fig:median_sersicn} but underdense-like
in Fig.~\ref{fig:median_dvcbtt}, partly due to redshift evolution.
However, when the overdensity (or underdensity) is prominent, we observe that the environment in
which a galaxy resides affects its morphology, so the appearance of the neutral bins does not
undermine the main conclusions of this paper.
}

To test the possibility that our choice of bins starting at $z=0.3$ with $\Delta z = 0.05$ is particularly unlucky in enhancing the effects we see, 
we carried out the same analysis using bins of the same width but shifted by $\Delta z/2$. We recomputed the overdensities keeping the parameters of the redshift distributions same as before.
Shifting the bins enhances the overdensities and underdensities in some cases and in other cases, it mixes the overdense and underdense regions to make them more neutral-like. 
The environmental trends still are observed in the enhanced bins and hence our conclusions are still applicable. 

\new{The number of galaxies in overdense bins is typically higher than the number in underdense bins by about a factor of 2. (See Table ~\ref{table:GalaxyCounts}). 
To eliminate the possibility (with great confidence) that this varying sample size is 
responsible for the trend observed, we repeated our analysis with only 50 per cent of overdense galaxies that were selected randomly. 
The vertical errorbars for overdense regions get bigger due to reduced sample size, but  the
statistical significance of our conclusions is still nearly as high as in the original analysis.}

\subsection{Mitigating the effects of line of sight fluctuations}\label{sec:mitigation}

\begin{figure}
 \centering
 \includegraphics[width=1.0\columnwidth]{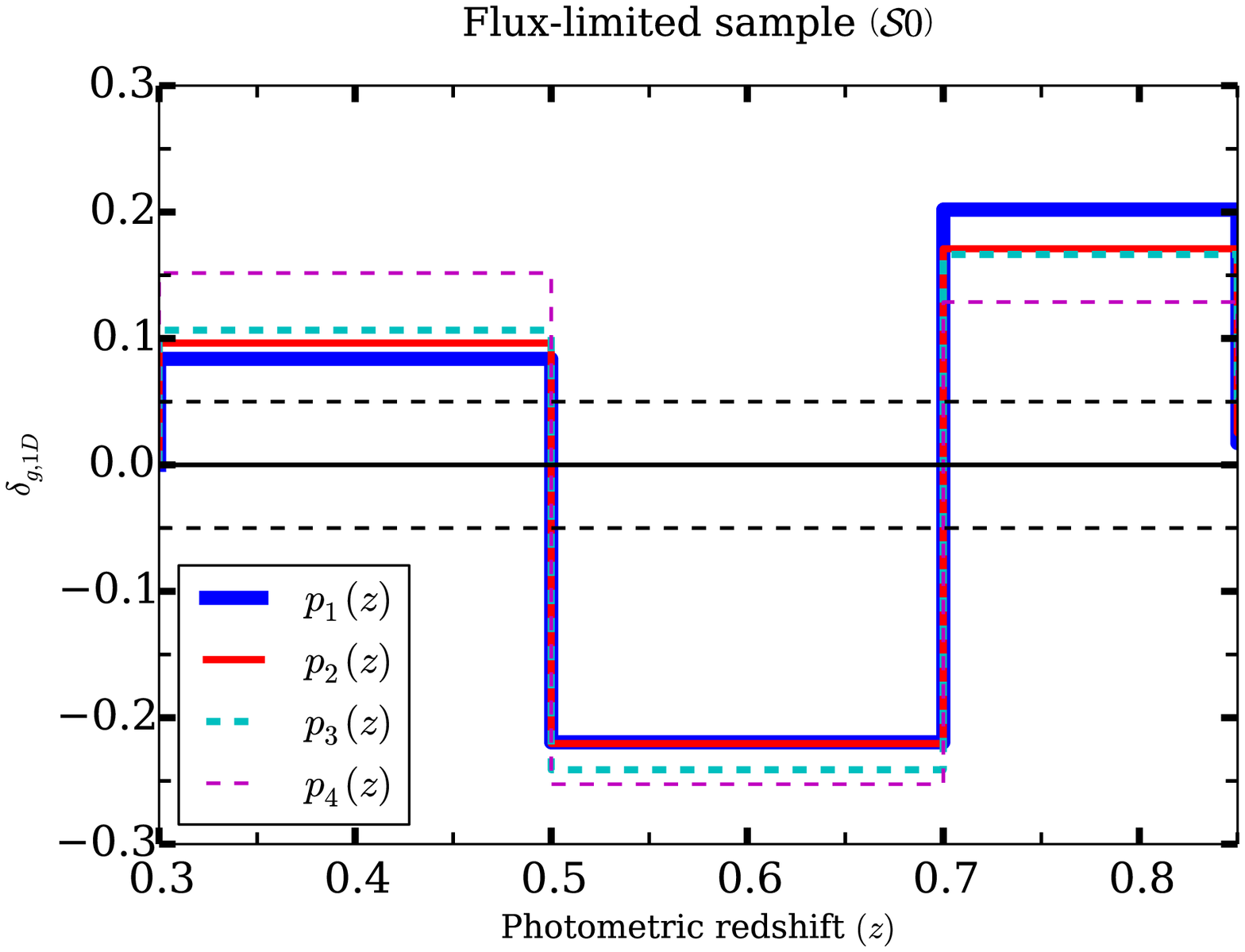}
 \includegraphics[width=1.0\columnwidth]{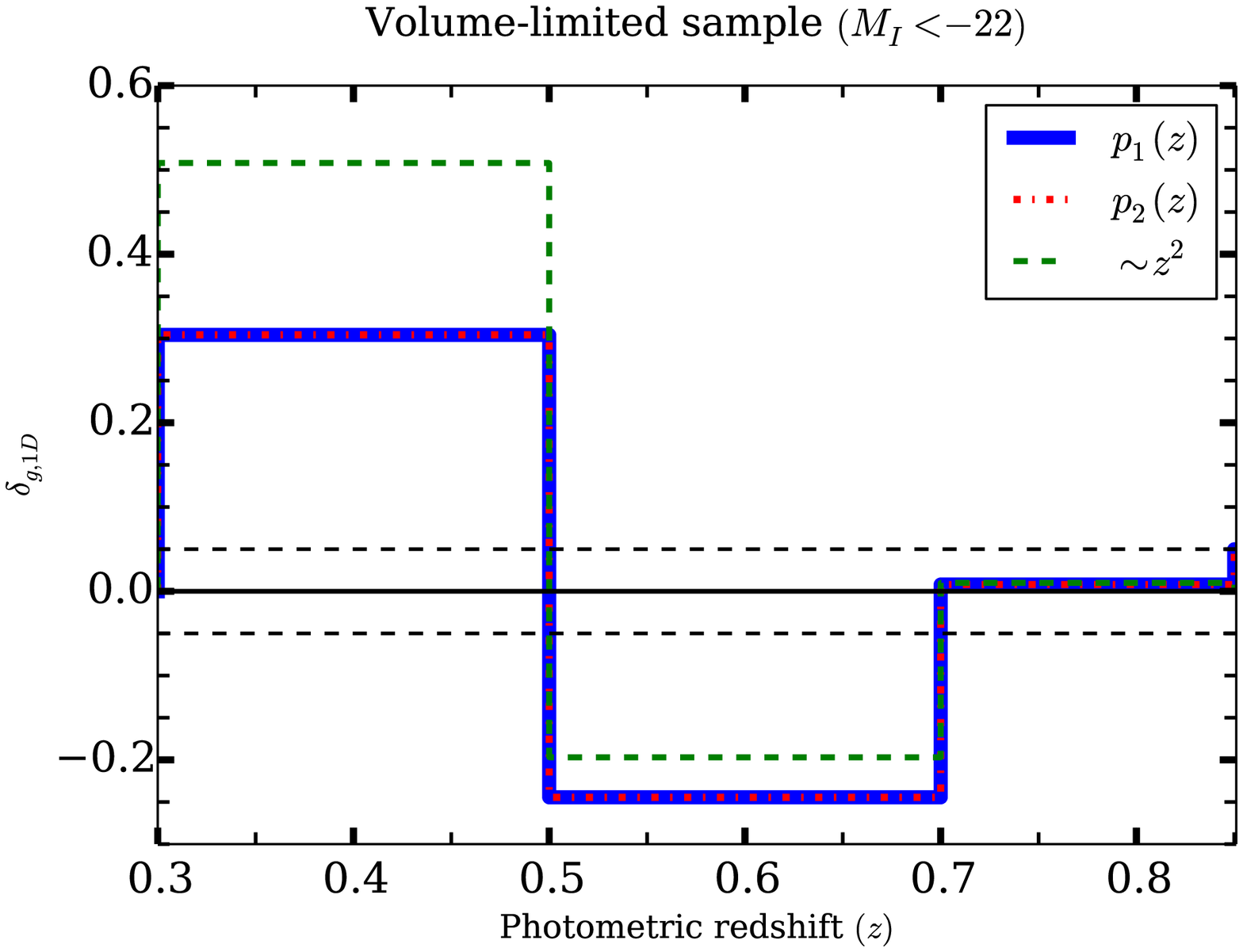}
 \caption{Plot of $\delta_{g,\text{1D}} =  N/N_{\text{mod}}-1$ with each functional form as the model
           for each of our new wide redshift bins discussed in Sec.~\ref{sec:mitigation}, for flux-limited (top) and
           volume-limited (bottom) samples.
          }
 \label{fig:redshift_wide}          
\end{figure}

\begin{figure}
 \centering
 \includegraphics[width=1.0\columnwidth]{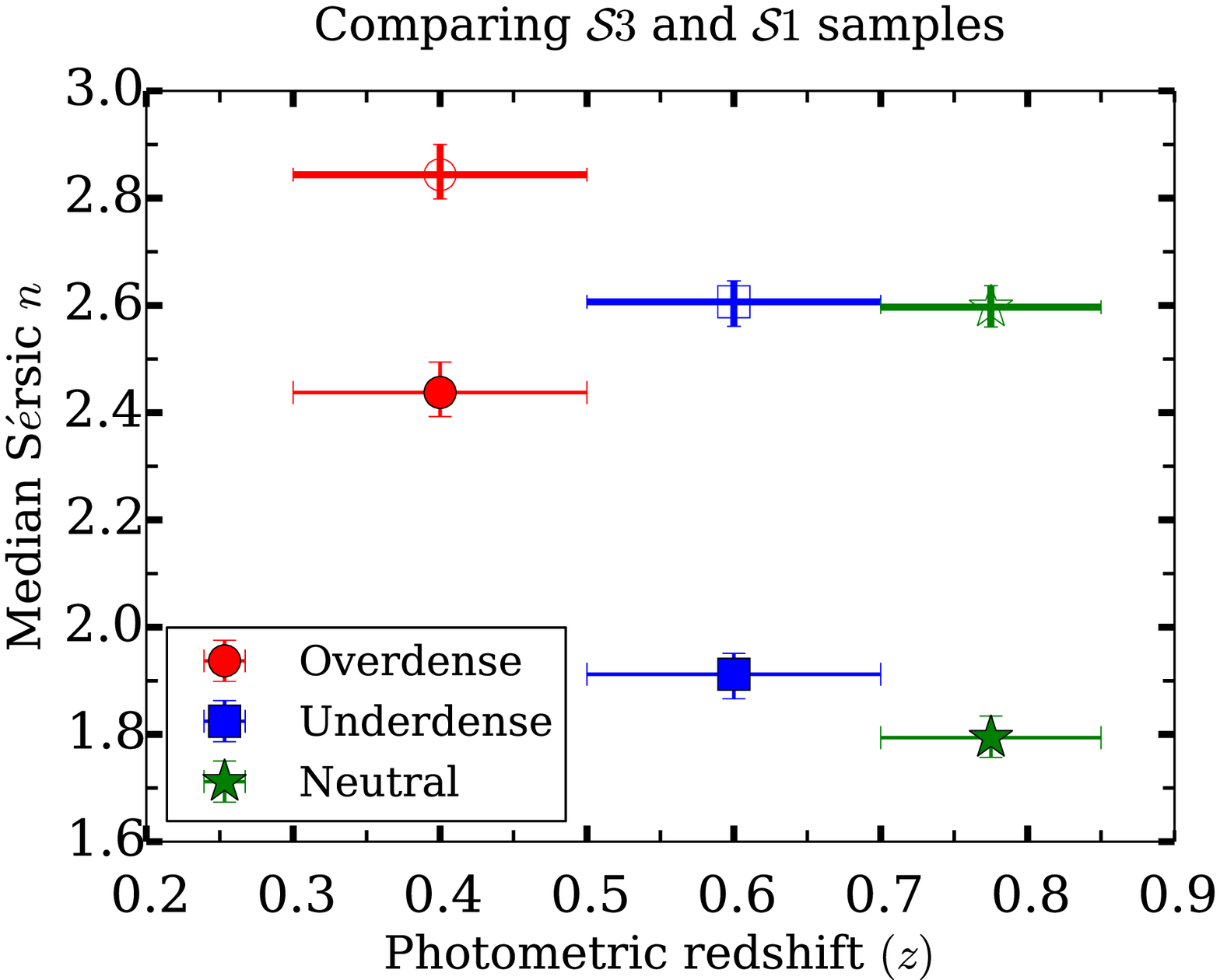} \
 \includegraphics[width=1.0\columnwidth]{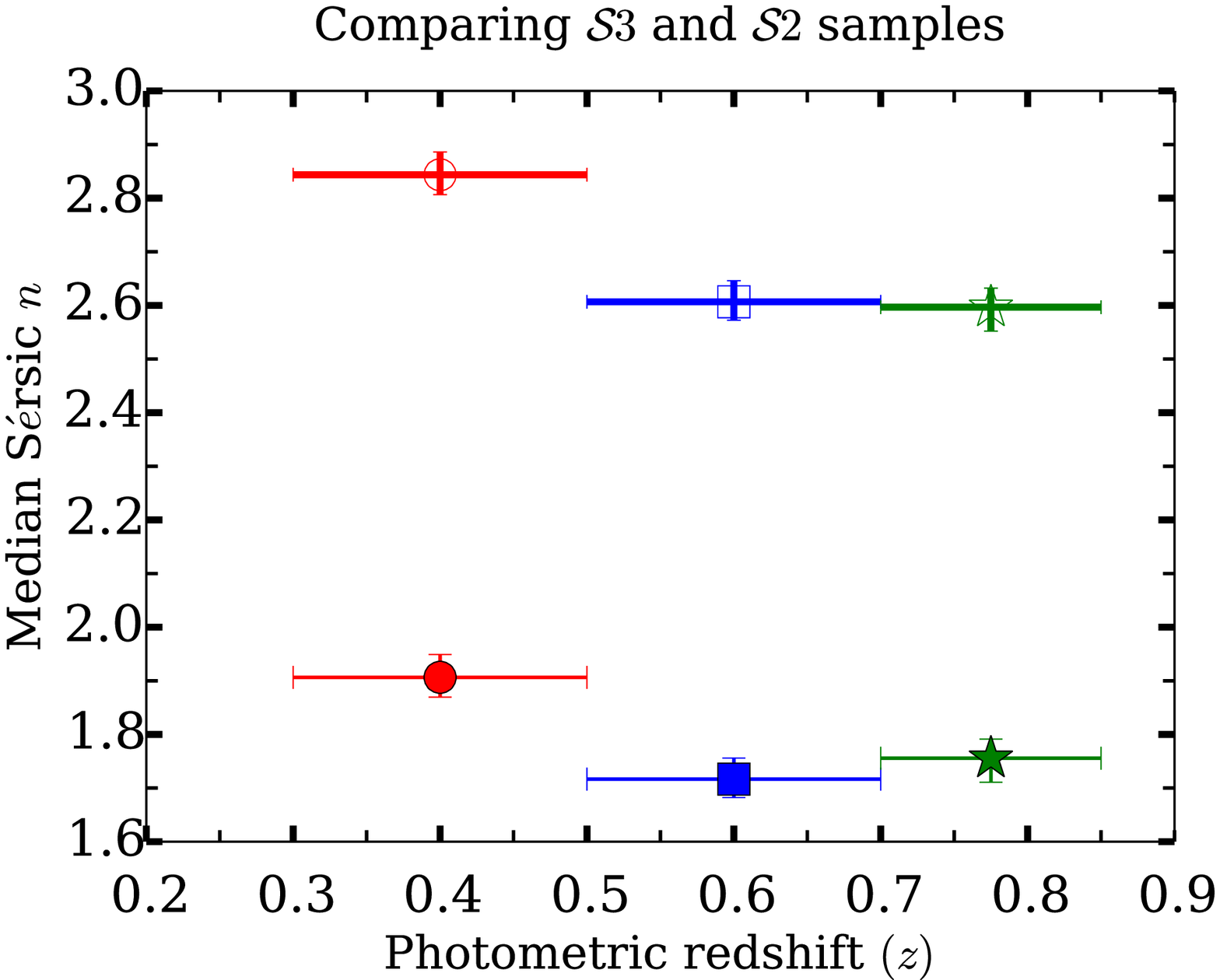} 
 \caption{ Median values of \sersic index, as a function of redshift
   for our wider redshift ranges used in Sec.~\ref{sec:mitigation}. The horizontal errorbars indicate
           the width of the redshift bin, while the vertical ones are $1\sigma$ errorbars obtained by
           bootstrapping. Points with open centers and thick errorbars correspond to the stellar-mass selected sample \s$3$
           and points with filled centers and thin errorbars correspond to the luminosity-selected samples \s$1$ and \s$2$.    
         }
 \label{fig:median_sersicn_wide}
\end{figure}

Since we have argued that cosmic variance is giving rise to
environmentally-based variations between galaxy populations in our
redshift slices that are $0.05$ wide, the natural question is how to
mitigate this effect so that it will not affect attempts to simulate a
realistic galaxy sample as a function of redshift.  The most obvious
approach is to repeat the analysis with wider redshift bins. Feigning ignorance of overdensities and underdensities along the line of sight, we choose the slices in redshift to be $\left[ 0.3 - 0.5 \right]$,
$\left[ 0.5 - 0.7\right]$ and $\left[ 0.7 - 0.85\right]$ (a nearly
even division of our entire redshift range) and redo the analysis,
beginning by checking the environmental classification for these wide bins.

Using the redshift distributions estimated earlier in this work (cf. Figs.~\ref{fig:redshift_fluxlimited}
and~\ref{fig:redshift_vollimited}), we obtain the overdensity estimates for
the wider bins, $\delta_{g,\text{1D}}$.
Although the $\left[ 0.7 - 0.85\right]$ bin seems to be overdense in
the top panel of Fig.~\ref{fig:redshift_wide}, it appears to be
environmentally neutral when we volume-limit the sample 
using the methods from Sec.~\ref{sub:volumelimiting}.
Since the latter is what we use to study the galaxy morphology, we
classify $\left[ 0.7 - 0.85\right]$ bin as `neutral'.
What is surprising is the fact that the lowest and middle redshift
slices still qualify as substantial overdensities and underdensities
despite our use of $\Delta z=0.2$.

As an example of what happens to morphological parameters, we show the
median \sersic index for three bins in
Fig.~\ref{fig:median_sersicn_wide}.  Comparing this with
Fig.~\ref{fig:median_sersicn}, we observe that the 
range of \sersic $n$ values has become smaller, as have the vertical errorbars, mainly due to the increase in the number of galaxies in each redshift bin. The results do not suggest that the values
are consistent across all redshift bins, and in particular, the
disparity between overdense and underdense regions is still quite
evident (with the same sign as before). However, in
Fig.~\ref{fig:median_sersicn}, the magnitude of that disparity between
overdensities and underdensities was around 20 per cent, whereas here
it is reduced to 7 per cent.  This seems to suggest that when we
choose wider redshift bins, some of the large-scale structure gets washed out, 
so the galaxy morphological parameters are less affected by cosmic
variance.  This is a promising result, which could be further improved
by either (a) non-blind selection of the wide redshift bins with respect to
known structure in the calibration fields, or (b) attempting some kind of
reweighting of the galaxy populations in redshift slices affected by
known structure.

\begin{figure}
 \centering
 \includegraphics[width=1.0\columnwidth]{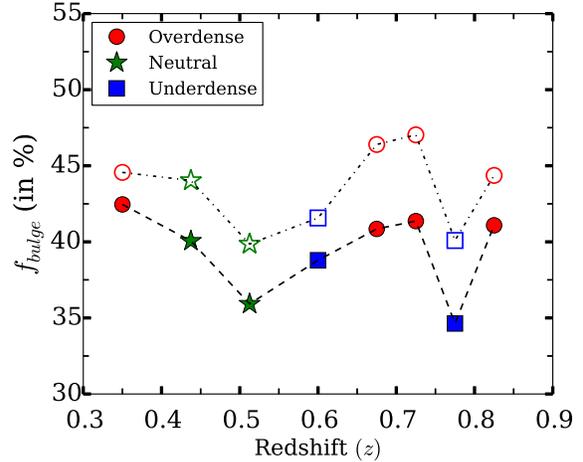} 
 \caption{The fraction of galaxies in \s1 (solid symbols) and
     \s3 (open symbols) that are classified as bulge-dominated as a
     function of redshift, including the correlation with redshift
     slices classified as overdense or underdense as indicated by the
     symbol shape and colour.}
 \label{fig:fbulge_z}
\end{figure}

As a test of the second option, we consider whether we can use some division of the
galaxies by colour or $B/T$ (as a proxy for colour) in order to
determine the structural parameters of galaxies as a function of
redshift for each of them separately.  If the effects we are seeing
can be cleanly described in terms of overdense (underdense) regions
being dominated by bulge- (disk-)dominated galaxies, then division
into samples based on $B/T$ may help reduce the effects that we have
described in previous subsections.  We test this idea by dividing the
sample into bulge- and disk-dominated galaxies based on requiring
$B/T\ge $ or $<0.5$.  The resulting bulge fraction $f_\text{bulge}$,
shown as a function of redshift in Fig.~\ref{fig:fbulge_z}, 
shows a mild effect from the large-scale structures in the COSMOS field.

\begin{figure}
 \centering
 \includegraphics[width=1.0\columnwidth]{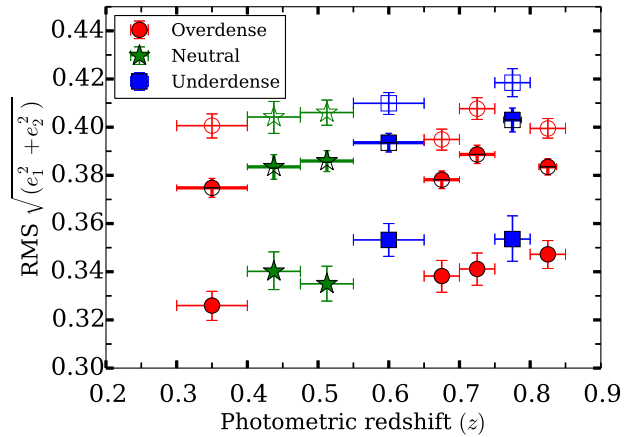} 
 \caption{The RMS distortion as a function of redshift for the
     full volume-limited galaxy sample \s3 (partially-full symbols), the
     disk-dominated galaxy sample (open symbols), and the
     bulge-dominated galaxy sample (full symbols).  The fact that the
     bulge-dominated sample has a lower RMS distortion than the
     disk-dominated sample is consistent with expectations.}
 \label{fig:rmsellip_bulge_disk}
\end{figure}

Fig.~\ref{fig:rmsellip_bulge_disk} shows the RMS distortion as
  a function of redshift for the full \s1 sample (as shown previously) and
  for the separate bulge- and disk-dominated samples.  As shown,
  the deviations in the values due to large-scale structure are
  clearly evident in all three cases, though slightly more prominent
  for the full sample and less so for the bulge-dominated sample.  For
comparison with the results for the full sample shown in
Table~\ref{table:pvalues_all}, we compute the KS (AD) test $p$-values
for consistency of axis ratios between overdense and underdense
regions for bulge- and disk-dominated samples.  For \s1, these
$p$-values are 0.16 (0.08) for bulge-dominated galaxies and 0.005
(0.001) for disk-dominated galaxies.  For \s3, these $p$-values are
0.03 (0.03) and $10^{-4}$ ($3\times 10^{-4}$), respectively.  With the
possible exception of bulge-dominated galaxies, it seems that
separation into two morphological samples is not enough to remove the
effect of the local environment on the ensemble properties like the
intrinsic ellipticity distribution.   Due to the limited sample size,
we did not explore the possibility of an even finer morphological
division of the sample, but with deeper {\em HST} datasets, this may
be a promising path to pursue.  However, if the nature of the bulges
or disks themselves is altered by the local environment, then division
purely into bulge- or disk-dominated sample will not be helpful in
mitigating environmental effects.

\section{Implications for future surveys}\label{S:implications}

As mentioned previously, {\em HST} data will be used by
current and future weak
lensing surveys to characterize the galaxy population in several
ways.  In this section, we estimate how the results shown in
Sec.~\ref{S:results} can affect estimates of shear calibration when
using {\em HST} to characterize the galaxy population.  We also
discuss the situations in which this is likely to be important for
current and future surveys from the space and ground.

\subsection{Magnitude of shear calibration bias}

Here we consider the impact of the findings in
  Sec.~\ref{S:results} on weak lensing shear calibration assuming that
  the COSMOS sample is used as a parent sample for simulations that
  are used to derive redshift-dependent shear calibrations.  This
  could be done either directly using the galaxy images themselves, or
  by fitting for parametric distributions of \sersic $n$, size, and
  shape in redshift slices, and then using those parametric
  distributions to make simulated images containing galaxies with
  \sersic light profiles that match those distributions.  We consider a few simple cases of how the
  above results affect shear calibration estimates.  It is likely that
  the answer to this question varies quite significantly with the type
  of shear estimation method used.  Some will be sensitive to the
  variations in morphology, others to the variation in the intrinsic
  ellipticity distribution, and many will be sensitive to both at some
  level.  We consider both of these issues in turn.

The intrinsic ellipticity distribution plays a role in nearly all
shear estimators.  In some, the role is explicit: for example, LensFit
\citep{2007MNRAS.382..315M,2008MNRAS.390..149K,2013MNRAS.429.2858M}
and the methods presented by \cite{2014MNRAS.438.1880B} require
accurate intrinsic ellipticity distributions as inputs, and
uncertainty in the distribution was one source of
systematic uncertainty in the CFHTLenS weak lensing results
\citep{2013MNRAS.432.2433H,2013MNRAS.429.2858M}.  The intrinsic
ellipticity distribution enters the calculation for other methods in
other ways.  For example, the re-Gaussianization method and several
other moment-based methods require calculation of a shear responsivity
\citep{BJ02,HS03} that describes how the galaxy population overall
responds to a shear, based on its intrinsic ellipticity distribution.
The responsivity can be calculated based on the observed shape
distribution, assuming that the uncertainties in the shears are known
well enough that their contribution to that distribution can be removed.
If a simulated sample in some redshift slice has a different intrinsic
ellipticity distribution and therefore responsivity, it could lead to
incorrect conclusions about shear calibration.  The responsivity
scales roughly like $1-e_\text{RMS}^2$, which means that deviations in
RMS distortion at the level of 0.01 due to local environments (Figs.~\ref{fig:rms_ellip} and~\ref{fig:rms_ellip_momentbased}) would become fractional shear
errors of
\begin{equation}
\frac{\Delta\gamma}{\gamma} \approx \frac{2 e_\text{RMS} \Delta
  e_\text{RMS} }{1-e_\text{RMS}^2} \approx 0.01.
\end{equation}
In the context of upcoming lensing surveys that seek to constrain
shears to better than the per cent level, a systematic error of this
magnitude in shear calibration is quite serious.

Regarding possible biases in the morphological mixtures of galaxies
due to overdensities or underdensities in the training sample, there
are results in the literature for several methods that show how shear
biases vary with morphology.  For example, for the maximum likelihood
fitting code {\sc im3shape}, figure 2 in \cite{2012MNRAS.427.2711K}
shows multiplicative biases for two-component \sersic profile galaxies
as a function of their bulge-to-total ratios (denoted there as
$F_b/(F_b+F_d)$, which we will equate with our $B/T$).  The shear
calibration bias scales roughly like $0.04 - 0.05(B/T)$ as $B/T$ goes
from $0$ to $1$.  Our results suggest that typical (median) $B/T$
values may be influenced by cosmic variance in the COSMOS field,
leading to fluctuations of order $0.05$.  The resulting variation in
the shear calibration  would therefore be $\sim 2.5\times 10^{-3}$, or
$0.25$ per cent shear calibration uncertainty.  For existing datasets
this is not very problematic, but for surveys like LSST, Euclid, and
WFIRST-AFTA, this would be a dominant part of the systematic error
budget.  As another example, for re-Gaussianization, figure 9 of
\cite{2012MNRAS.420.1518M} shows that as \sersic $n$ goes from $1$ to
$6$, the shear calibration bias varies by $2$ per cent.  In this case,
since we have shown that the median value of \sersic $n$ can vary by
$\sim 0.4$ due to morphology-density correlations, this suggests that
the shear calibration for re-Gaussianization could be misestimated by
$\sim 2\times 10^{-3}$, or 0.2 per cent.  This too is acceptable in
existing datasets, but not those that will be used for shear
estimation in the next decade.

The estimates in this subsection are rough illustrations of
  the magnitudes of these effects.  Other aspects of shear calibration
  that could be affected relate to the use of {\em HST} data to
  estimate the impact of detailed galaxy morphology, or to calibrate
  the effect of colour gradients
  \citep{2012MNRAS.421.1385V,2013MNRAS.432.2385S}.  In the latter
  case, what is most relevant in this work is our finding that $B/T$
  exhibits environmental dependence, which likely translates into
  environmental dependence of colour gradients.  Unfortunately, we
  cannot directly test the strength of any colour gradient variations
  with environment using COSMOS data, due to the fact that there is
  only single-band coverage in much of its area.  We note that our
  findings may seem to be at odds with the conclusions in
  \cite{2013MNRAS.432.2385S} based on synthetic galaxy models that the
  existing area of {\em HST} coverage with $\ge 2$ bands is sufficient
  for calibration of colour
  gradients\footnote{\cite{2012MNRAS.421.1385V} showed that the
    fluctuations in colour gradients within the source galaxy sample
    in the Euclid weak lensing survey due to environmental effects
    will not be a significant source of uncertainty.  However, the
    question addressed in that work is different from the question
    considered here, which is the impact of fluctuations in colour
    gradients in the training sample used to derive the corrections,
    rather than in the galaxy sample to which those corrections will
    be applied.}.  This is particularly striking given that, as shown
  there, the dominant galaxy sample with $2$-band coverage is the
  AEGIS dataset, which has substantially smaller area than COSMOS, and
  therefore should exhibit a stronger influence of cosmic variance.
  The presence of other fields besides AEGIS (e.g., GOODS)
    should help mitigate this effect given that the large-scale
    structure in the two fields will be completely uncorrelated, but
    the number of fields is still not large and the area is dominated
    by a small number of them\footnote{For context,
        \cite{2015APh....63...81N} 
        showed that for tests of photometric redshift quality, the
        number of spectroscopic fields of this size that would be
        required for future surveys to avoid the influence of cosmic
        variance ranges from many tens to hundreds.}.

However, it is important to bear in mind that the method
  proposed in \cite{2013MNRAS.432.2385S} involves determining not just
a redshift-dependent correction but also a type-dependent colour
gradient correction, which could partially mitigate the effects of the
environment dependence seen here.  Moreover, the colour gradient
effect is higher order than the intrinsic ellipticity distribution,
and thus may be less susceptible to systematics due to the
environmental effects considered here.   We showed in
Sec.~\ref{sec:mitigation} that a simple type-dependent split does not
remove the effects of environment on the intrinsic ellipticity
distribution, but testing whether it (or a more complex type or colour
split) is
enough to remove the effects of colour gradients in Euclid is a more
complicated analysis that is beyond the scope of this work. It seems that a future study using
  an {\em HST} field with more than one band would be warranted, to
  test whether or not our results suggest a discrepancy with those
    of 
  \cite{2013MNRAS.432.2385S}.  It is possible that the connection between
  colour gradients and $B/T$ is weak enough that our results do not imply a
  problem with colour gradient calibration, or that 
    division into several galaxy types or colours is truly enough to remove the colour
    gradient effect due to its being a higher order effect, consistent with the findings of 
    \cite{2013MNRAS.432.2385S} using synthetic galaxy models.

\subsection{Effective impact on current and future surveys}

In addition to the order of magnitude of these effects
  presented in the previous subsection, it is important to bear in
  mind how current and future surveys plan to use {\em HST} data.

For example, before the Euclid survey begins, when carrying
  out tests of shear estimation methods, their simulated data will be
  based on {\em HST} in many ways: estimation of simple aspects of
  morphology (\sersic $n$, bulge fraction), intrinsic ellipticity
  distribution, colour gradient calibration, and higher order moments
  (detailed morphology).  Our results suggest that care should be
  taken to ensure that those simulations are not overly influenced by
  environment effects in the {\em HST} data, so that incorrect
  conclusions will not be drawn about the redshift-dependence of shear
  calibration for shear estimation methods to be used by the survey.

However, once the Euclid survey is
  underway, the derivation of simple galaxy morphology and the
  intrinsic ellipticity distribution will be based on the 40~deg$^2$
  Euclid deep field, which goes two magnitudes fainter than the rest of
  the survey.  With that data in hand, the Euclid weak lensing
  results will be less reliant on the much smaller and more
  cosmic variance-limited {\em HST} fields, using them only for colour
  gradient calibration and estimates of the impact of detailed galaxy
  morphology (since the {\em HST} resolution is higher than that
  of Euclid). The findings of \cite{2014MNRAS.439.1909V} demonstrate that the area of
  this deep field is sufficient to accomplish the goal of determining
  the intrinsic ellipticity distribution at the accuracy required for
  shear calibration purposes for the Euclid survey.

In contrast, for ground-based surveys such
  as LSST, high-resolution space data will play a more important role
  in the understanding of the galaxy 
  intrinsic ellipticity and morphology distributions, since even a deep
  field in a ground-based survey faces fundamental resolution limits
  that prevent the derivation of detailed information about the faint
  galaxy population.  Galaxies that are near the resolution limit for
  a ground-based survey are still very well-resolved in {\em HST},
  making it the best resource for detailed information about them.
  Once the Euclid deep field is publicly available, the information
  from it will be beneficial to ground-based surveys as well.

Finally, our results for the effect on the intrinsic
  ellipticity distributions suggest that even current surveys should
  be careful to avoid this effect.  While use of COSMOS without
  accounting for the effect that we have identified will cause a bias
  that is similar to the final requirements on the shear systematic
  errors for Stage III surveys, there are other elements in the
  systematic error budget, so ideally this effect should be mitigated
  somewhat using, for example, the wider redshift binning strategy
  that we tested in Sec.~\ref{sec:mitigation}.  Use of significantly smaller fields
  than COSMOS, while possibly helpful in building up a deeper galaxy
  sample, should naturally increase the impact of cosmic
  variance on the training dataset, which should be quantified as part
  of the systematic error budget.

\section{Conclusions}
\label{S:summary}

In this study, we have shown that the shape distributions of galaxies
(to a statistically significant degree) and morphological parameters
like \sersic $n$ and bulge-to-total ratios (more marginally) depend on
the local environments when dividing up the COSMOS sample into
redshift slices along the line of sight.  The redshift slices used for
our primary analysis had
a width of $\Delta z=0.05$.  Our findings are robust to the choice of
shape estimator from \sersic profile fits vs.\ using centrally weighted
moments-based shear estimates.

These findings are relevant to attempts to use {\em HST}-based galaxy
samples to calibrate shear estimates in weak lensing surveys.  In
general, the approach would be to define galaxy samples using all
galaxies in redshift slices, and determine a redshift-dependent shear
calibration.  Our findings highlight the danger in such an approach:
while we would like our simulations to include true evolution in
galaxy properties with redshift, this approach also includes spurious
variations in galaxy properties due to the large-scale structure
within the COSMOS field.  Since the fidelity of weak lensing shear
estimates depends sensitively on the intrinsic shape distribution and
galaxy morphological parameters, the conclusions for the
redshift-dependent shear calibration would be incorrect.  As shown in
Sec.~\ref{sec:mitigation}, these errors
are reduced as the redshift slices that are used become wider, so
that the impact of local overdensities becomes washed out.  However,
our results suggest that even $\Delta z=0.2$ may not be wide enough
(and this is becoming dangerously close to the size of tomographic
redshift bins to be used for weak lensing analysis in upcoming
surveys).  Thus, more complex schemes may become necessary to fully
overcome this issue, depending on exactly how the {\em HST}
  data is to be used.  As discussed in Sec.~\ref{S:implications},
  particular care may be needed in ground-based surveys that will use
  the {\em HST} data to model many aspects of the galaxy population.
  In contrast,  for Euclid, use of the relatively large area Euclid deep
  data to constrain many aspects of the galaxy population means that
  these issues with {\em HST} data are less important, though still
  not completely ignorable.

It is important to keep in mind the nature of COSMOS with respect to
other possible {\em HST} training samples.  COSMOS represents the
largest contiguous field surveyed by the {\em HST}, with the sizes of
other major {\em HST} fields such as GOODS and AEGIS lagging by at least a
factor of 9 (for AEGIS, and more for GOODS).  Combinations
  of CANDELS with existing datasets may ultimately be as large as
  $1/6$ the COSMOS area.
Hence, if cosmic variance due to structures along 
the line of sight in COSMOS are problematic for its use as a training
sample for weak lensing simulations, studies that use even smaller
area training samples are even more prone to errors, with the UDF
serving as an extreme case.  If the size of the {\em HST} survey is
small enough, there is no reason {\em a priori} to suppose that the
galaxy population is typical even when using all galaxies along the
line of sight, without any division into redshift bins.  Of course,
future surveys are unlikely to pick just a single survey to serve as
the basis for their image simulation training sample, but rather
will combine as many as possible.  Combining multiple surveys will
reduce the cosmic variance and therefore the significance of the
effects discussed in this work.
However, as COSMOS is significantly larger than other \emph{HST} surveys, combining COSMOS with smaller fields is unlikely to ameliorate this effect.
Thus, it will be important to carefully choose the size of redshift slices used to
derive properties of the galaxy population so as to be minimally
affected by this issue.

A final consideration is the question of how applicable these results
using volume-limited samples are to simulations of upcoming weak
lensing surveys, which will exclusively use flux-limited samples.  For
our analysis, the volume-limiting sample was necessary to avoid
complications due to varying galaxy populations in each redshift
slice, allowing us to isolate purely environmental effects.  In
principle, if the morphology-density correlations that we have
identified turn out to not exist for intrinsically fainter galaxy
populations, then at low redshift (where a flux limited sample will
include galaxies that are intrinsically much fainter than at high
redshift), the effects will be less serious for upcoming lensing
surveys.  However, we do not have any particular reason to believe
that these effects will vanish for fainter galaxies.  Moreover, at
higher redshift where only intrinsically bright galaxies can be seen,
the effect should be present at a level similar to what we have found
here.  Since higher redshift galaxies tend to dominate cosmological
shear estimates (due to their higher shears), our findings will be
important to take into account.  It would also be advisable to
  carry out a future study of this effect at higher redshift (beyond
  $0.85$), using for example the data from the CANDELS survey.

In conclusion, our results have serious implications for the plans to
create realistic image simulations that will be used to derive
redshift-dependent shear calibrations for upcoming weak lensing
surveys.  If care is not taken to mitigate this effect, then the
cosmic variance in the training sample may bias the conclusions
regarding shear calibration for redshift slices that represent
significant overdensities or underdensities compared to the typical
galaxy population.  This is particularly a problem when using the
smaller {\em HST} surveys, where a single galaxy cluster or a void
could completely dominate the galaxy population in a given redshift
slice.  To mitigate this problem, it will be imperative to (a) collect
training data from widely separated patches on the sky, and (b) take
care to use redshift slices that are broad enough that these effects
are reduced, so as to wash out the effect of any signal overdensity or
underdensity on the simulated galaxy population.  By employing these
mitigation schemes, there is every reason to believe that the effect
we have identified can be reduced to a small component of the
systematic error budget of major upcoming lensing surveys.

\section*{Acknowledgments}

The authors would like to thank Thomas Kitching, Henk
  Hoekstra, Tim Schrabback \new{and the anonymous referee} for their helpful comments on this work.
AK and RM acknowledge the support of NASA ROSES 12-EUCLID12-0004, and
program HST-AR-12857.01-A, provided by NASA through a grant from the
Space Telescope Science Institute, which is operated by the
Association of Universities for Research in Astronomy, Incorporated,
under NASA contract NAS5-26555. RM acknowledges the support of an Alfred P. Sloan Research Fellowship. 
Kavli IPMU is supported by World Premier International Research Center Initiative (WPI), MEXT, Japan.
We thank Alexie Leauthaud for 
many useful discussions.

\bibliography{ref}
\end{document}